\documentclass[a4paper,usenatbib]{mn2e}
\usepackage{float}
\usepackage{times}
\usepackage{caption}
\usepackage{subfigure}
 \usepackage{graphicx}
 \usepackage{graphics}
 \usepackage{amsmath}
 \usepackage{lipsum} 
 \usepackage{afterpage}
 \newcommand{\hi}{\mbox{H{\sc i}}}
 \newcommand{\aap}{A\&A}

\newcommand{\aj}{AJ}
\newcommand{\apj}{ApJ}

\newcommand{\apjl}{ApJL}

\newcommand{\mnras}{MNRAS}
\newcommand{\araa}{ARA\&A}
 \renewcommand\appendix{\par
  \setcounter{section}{0}
  \setcounter{subsection}{0}
  \setcounter{figure}{0}
  \setcounter{table}{0}
  \renewcommand\thesection{ \Alph{section}}
  \renewcommand\thefigure{\Alph{section}\arabic{figure}}
  \renewcommand\thetable{\Alph{section}\arabic{table}}
}
\title[Galaxy Mass Models: MOND vs Dark Matter halos]{Galaxy Mass Models: MOND versus Dark Matter Halos}
\author[T. Randriamampandry \& C. Carignan]{Toky Randriamampandry$^{1}$\thanks{tokyr@ast.uct.ac.za} and Claude Carignan$^{1}$\thanks{ccarignan@ast.uct.ac.za}\\
$^{1}$ Department of Astronomy, University of Cape Town, Private Bag X3, Rondebosch 7701, South Africa}

\begin{document}

\date{}

\pagerange{\pageref{firstpage}--\pageref{lastpage}} \pubyear{2013}

\maketitle
\label{firstpage}

\begin{abstract}

Mass models of 15 nearby dwarf and spiral galaxies are presented. The galaxies are selected to be homogeneous in terms of the method used to determine their distances, the sampling of their rotation curves (RCs) and the mass--to--light ratio (M/L) of their stellar contributions, which will minimize the uncertainties on the mass model results. Those RCs are modeled using the MOdified Newtonian Dynamics (MOND) prescription and the observationally motivated pseudo-isothermal (ISO)  dark matter (DM) halo density distribution.  For the MOND models with fixed (M/L), better fits are obtained when the constant a$_{0}$ is allowed to vary, giving a mean value of (1.13 $\pm$ 0.50) $\times$ 10$^{-8}$ cm s$^{-2}$, compared to the standard value of 1.21 $\times$ 10$^{-8}$ cm s$^{-2}$. Even with  a$_{0}$ as a free parameter, MOND provides acceptable fits (reduced $\chi^{2}_{r}$  $<$ 2) for only 60 \% (9/15) of the sample. The data suggest that galaxies with higher central surface brightnesses tend to favor higher values of the constant a$_{0}$. This poses a serious challenge to MOND since a$_{0}$ should be a universal constant. For the DM models, our results confirm that the DM halo surface density of ISO models is nearly constant at  $ \rho_{0} \ R_{C} \sim 120\ M_{\odot} \ pc^{-2}$. This means that if the (M/L) is determined by stellar population models, ISO DM models are left with only one free parameter, the DM halo central surface density.

\end{abstract}

\begin{keywords}
Cosmology: dark matter; galaxies: kinematics and dynamics
\end{keywords}

\section{Introduction}

Since the 1970's, it is well known that there is a discrepancy between the visible mass and the dynamical mass of galaxies (e.g. \citealt{1970ApJ...160..811F,1973A&A....24..411S, 1975ApJ...201..327R, 1978PhDT.......195B}). The commonly accepted explanation of the galaxy mass discrepancy is to assume a more or less spherical halo of unseen  Òdark matterÓ in addition to the visible baryonic mass in the form of stars and gas (see e.g: \citealt{1985ApJ...294..494C}). 
The distribution of dark matter in galaxies can be defined by a theoretical or an empirical density distribution profile.
Several density profile models for the distribution of dark matter (DM) are presented in the literature.
The two most commonly used models are (see also e. g:  \citealt{1969AN....291...97E, 1995ApJ...447L..25B, 1998ApJ...502...48K}):
\begin{itemize}
  \item the pseudo-isothermal (ISO) dark matter halo model (observationally motivated model)
  \item the Navarro-Frenk-White (NFW) dark matter halo model \citep{1997ApJ...490..493N} (derived from $\Lambda$CDM N-body simulations)
\end{itemize}
The highest quality H{\sc i} RC
s (RCs) available to date are those for the THINGS (The \hi\ Nearby Galaxies Survey) sample \citep{Walter:2008bs}.  For that sample, the RCs of 19 galaxies were derived by \cite{de-Blok:2008oq} and their mass distributions were modeled using the ISO and the NFW halo density profiles. \cite{de-Blok:2008oq} found that most of the galaxies in their sample preferred the observationally motivated core-dominated ISO halo over the cuspy NFW halo. This is known as the core-cusp controversy (see  \citealt{2010AdAst2010E...5D} for a review). This is why the ISO models will be used in this paper for the DM models.

An alternative to the missing mass problem is the MOdified Newtonian Dynamics (MOND) \citep{Milgrom:1983rr}. Milgrom postulates that at small accelerations the usual Newtonian dynamics break down and that the law of gravity needs to be modified. MOND claims to be able to explain the mass discrepancies in galaxies without the need for dark matter but with the introduction of a universal constant a$_{0}$, which has the dimension of an acceleration.

The phenomenological success of MOND to reproduce the observed RCs  of galaxies has attracted a huge interest in the astronomy community for the last three decades. One of the first studies was made by  \cite{1987AJ.....93..816K} using \hi\ RCs of spiral galaxies and his conclusions were not favorable to MOND. Kent's work was criticized by \cite{1988ApJ...333..689M} who pointed out  possible errors in the distances and inclinations used and the possibility that not all the H{\sc i} gas had been detected. 

Dwarf galaxies are critical to test a theory such as MOND because these objects present the largest discrepancies between the visible mass and the dynamic mass and their small accelerations put them mostly into the MOND regime. \cite{1989ApJ...345L..17L} concluded that MOND could not reproduce the observed RCs of a sample of dwarf galaxies where the luminous mass was dominated by the gas and not the stars. However,  \cite{1991ApJ...367..490M} disapproved Lake's conclusions arguing again that there could be large errors in the distances and inclinations. 

 \cite{1991MNRAS.249..523B} were the first to estimate the value of a$_{0}$ by fitting the RCs of bright spiral galaxies.  The MOND parameter  a$_{0}$ was taken as a free parameter during their fitting procedure. They found an average value of  a$_{0}$ = 1.21 x 10$^{-8}$ cm s$^{-2}$.  However, the distances used were quite uncertain since Hubble's Law was adopted as the distance indicator for most of the  galaxies in their sample. The \cite{1991MNRAS.249..523B} results  were confirmed by \citet{1996ApJ...473..117S}  and \cite{1998ApJ...503...97S} using the same method.  \citet{1996ApJ...473..117S} used a larger sample of 22 galaxies selected from the literature but only M33 and NGC 300 had Cepheid distances. Since then, the value found by  \cite{1991MNRAS.249..523B} is considered  as the standard value for a$_{0}$. \cite{Bottema:2002kl} were the first to use a sample with Cepheid-based distances only and found that a lower value for  a$_{0}$  (0.9 x 10$^{-8}$ cm s$^{-2}$) yields better fits to the RCs. A complete review about the previous tests of MOND is presented in \cite{Sanders:2002qe}. In that review the basic framework of MOND is explained.  RCs fits using MOND prior to 2002 are presented and the different capabilities of MOND are listed. The most recent review on MOND and its implications for cosmology can be found in \cite{2012LRR....15...10F}. 
 
The analysis  by \cite{1998ApJ...508..132D} was the first to used LSB galaxies in the context of MOND. Those systems are good candidates to test MOND because their accelerations fall below the MOND acceleration limit a$_{0}$. \cite{1998ApJ...508..132D} used a sample of 15 LSB galaxies. They  found that MOND is successful to reproduce the shape of the observed RCs for three quarters of the galaxies in the sample. The most recent study of LSB galaxies in the context of MOND was done by \cite{Swaters:2010hc}.  MOND produced acceptable fits for also three quarters of the galaxies in their sample. The correlation between a$_{0}$ and the extrapolated central surface brightness of the stellar disk was also investigated. \cite{Swaters:2010hc} found that there might be a weak correlation between a$_{0}$ and the R-band central surface brightness of the disk. This is shown in their Figure 7. Their interpretation was that galaxies with lower central surface brightness had lower values for a$_{0}$. Such a correlation would be in contradiction with MOND since a$_{0}$ should be an universal constant.
The reliability of the MOND mass models depends strongly on the sensitivity and spatial resolution of the observed RCs. More extended RCs are needed to trace the matter up to the edge of the galaxies but higher resolution is required in the inner parts.

RCs derived from the THINGS sample satisfy these criteria. THINGS consists of 34 dwarfs and spiral galaxies observed in \hi\ with the Very Large Array in the B, C and D configurations \citep{Walter:2008bs}. \cite{Gentile:2011th} used a subsample of 12 galaxies from \cite{de-Blok:2008oq} and modeled their mass distributions using the MOND formalism. 
They performed one and two parameter MOND fits and recalculated the value of a$_{0}$ which turned out to be similar to the standard value estimated in  \cite{1991MNRAS.249..523B}. Their average value for a$_{0}$ is 1.22 x 10$^{-8}$ cm s$^{-2}$ using the simple $ \mu$-function \citep{2006ApJ...638L...9Z} for the interpolation between the Newtonian and the MONDian regimes.
They also looked at the correlation between a$_{0}$ and central surface brightness in the 3.6 micron band and found that there was no correlation. However their points have a large scatter and they used the bulge central surface brightness  instead of the extrapolated disk central surface brightness. 

Slowly rotating gas rich galaxies are good candidates to test MOND because their accelerations are below a$_{0}$ and the baryonic mass is dominated by gas and not stars. \cite{2013AJ....145...61S} studied a sample of five such galaxies and found significant departures between the observed RCs and the ones predicted by MOND (see also \citealt{2012MNRAS.426..751F}). Recently, \cite{2013AJ....146...48C} presented a detailed study of the magellanic--type spiral galaxy NGC 3109 using VLA H{\sc i} data from \cite{1990AJ....100..648J} and new data from the SKA pathfinder KAT-7 telescope in the Karoo desert in South Africa. With both data sets, they found that MOND cannot fit the observed RC of NGC 3109, while their distance and inclination are well determined. On the other hand, the ISO dark matter halo model gives a very good fit to both data sets. 
 
A sample of fifteen galaxies, mostly with Cepheid-based distances, will be used for the present study with the aim to minimize the errors coming from the adopted distance. Some RCs have been resampled to have a homogeneous sample of RCs with independent velocity points in order to give significance to the goodness of the fits of the mass models. The (M/L)s used have also been determined in a homogeneous way using stellar population models predictions, instead of leaving them as free parameters.

This paper is organized as follows: 
the sample selection is presented in section 2, the methods used for the mass models are explained in section 3, results are shown in section 4, followed by the discussion in section 5 and the conclusions in section 6.

\section[]{Sample}
\begin{table*}
\footnotesize
 \centering
 \begin{minipage}{140mm}
\caption{Properties of the galaxies in the sample}
\begin{center}
\begin{tabular}{c  c   c   c   c   c    c  c }
\hline\hline

Name& P. A. & Incl.  & Distance &Method[Ref] & m$_{B}$  & M$_{B}$ & Type \\
 &  $^{\circ}$ & $^{\circ}$ & Mpc & & mag & mag & \\
 1 & 2 & 3 & 4 & 5 & 6 & 7 & 8\\ 
\hline
DDO 154 &230&66&4.30 $\pm$ 0.54&bs[K04]& 13.94 & -14.23 & IB(s)m\\\\
IC 2574  &  53.4 &   55.7  &4.02 $\pm$ 0.41&rgb[K04]&10.80 & -17.21 &SAB(s)m\\\\ 
NGC 0055 & 109.7 & 76.9 & 1.94 $\pm$ 0.03 & cep[G08]& 9.60& -16.79&SB(s)m \\\\
NGC 0247 & 170.0 & 74.0 & 3.41 $\pm$ 0.17 & cep[G09] & 9.70&-17.95 &SB(s)cd \\\\
NGC 0300 & 310.5& 42.3&1.99 $\pm$ 0.04&cep[G05]& 8.72 & -17.67  &SA(s)d\\\\
NGC 0925 &286.6 &66.0 & 9.16 $\pm$ 0.63&cep[F01]& 10.69 & -19.13 & SAB(s)d\\\\
NGC 2366 &39.8 &63.8 &3.44 $\pm$ 0.31& cep[T95] & 11.53 & -16.13  &IB(s)m \\\\
NGC 2403 & 123.7 & 62.9  & 3.22 $\pm$ 0.14&cep[F01] & 8.93 & -18.60  &SAB(s)cd\\\\
NGC 2841 & 152.6& 73.7& 14.10 $\pm$ 1.50&cep[F01]  &10.09 & -20.66&SA(r)b \\\\
NGC 3031 & 330.2 & 59.0  & 3.63 $\pm$ 0.25&cep[F01]&7.89 & -19.89 &SA(s)ab\\\\
NGC 3109 &93.0 & 75.0& 1.30 $\pm$ 0.02&cep[S06] & 10.39 & -15.18 & SB(s)m\\\\
NGC 3198 & 215.0 & 71.5  & 13.80 $\pm$ 0.95&cep[F01] & 10.87 & -19.83&SB(rs)d\\\\

NGC 3621 &  345.4 & 64.7  & 6.64 $\pm$ 0.46&cep[F01]& 10.28 & -18.82 &SA(s)d\\\\
NGC 7331 & 167.7 & 75.8 & 14.72 $\pm$ 1.02&cep[F01]& 10.35 & -20.49 &SA(s)b\\\\
NGC 7793 &290 &50 & 3.43 $\pm$ 0.10& cep[P10] &9.17 &-18.79 &SA(s)d\\
\hline
\end{tabular}
\end{center}
\addtocounter{footnote}{-2}
\small{bs: brightest stars, rgb: red giant branch, cep: cepheid variables; F01: \cite{Freedman:2001eu}; S06: \cite{2006ApJ...648..375S}; K04: \cite{Karachentsev:2004bh}; T95: \cite{1995AJ....110.1640T}; P10: \cite{2010AJ....140.1475P}; G05: \cite{2005ApJ...628..695G}; G08: \cite{2008ApJ...672..266G}; G09: \cite{2009ApJ...700.1141G}}\\
\small{col. 1: Galaxy name; col. 2: Position Angle; col. 3: Inclination; col. 4: Distance with the uncertainty; col. 5: Method used to measure the distance \& reference; col. 6: Apparent magnitude taken from the RC3 catalog; col. 7: Absolute magnitude; col 8: Morphology type.\\

}
\label{prop}
\end{minipage}
\end{table*}

The two main selection criteria for the final sample of 15 dwarf and spiral galaxies are the method used to determine their distance and the availability of high quality \hi\ RCs in terms of spatial resolution and sensitivity. Cepheid based distances, using their period--luminosity relation,  are commonly considered as the most accurate for nearby galaxies. We were able to use this method for all but two of the galaxies in the sample. The method used to measure the distance  and references are given in column 5 of table 1. 

Another criteria is to have a sample of galaxies spanning a wide range of luminosities and morphological types from dwarf irregular to bright spiral galaxies. Therefore DDO 154 and IC 2574 are included  even if Cepheid distances are not available. Furthermore, these two galaxies are gas dominated galaxies and exhibit large discrepancies between their visible mass and their dynamical mass which makes them ideal objects to test MOND and DM halo models. 

Eleven galaxies of the final sample are part of THINGS, their RCs being derived by \cite{de-Blok:2008oq}. For the others, the RCs and gas distributions are taken from \cite{2013AJ....146...48C} for NGC 3109, from  \cite{2011MNRAS.410.2217W} for NGC 300, from \cite{1991AJ....101..447P} for NGC 55 and from \cite{1990AJ....100..641C} for NGC 247. 
There is considerable overlap between the sample used in this work and \cite{Gentile:2011th}, 7 out of 15 galaxies are common to both studies. However, IC 2574, NGC 925 and NGC 2366 are included in this study, which were omitted by \cite{Gentile:2011th}
 because of the presence of holes and shells and non-circular motions. The RCs of these galaxies were derived using the bulk velocity fields only (Oh et al. 2008) which remove the effect of holes and shells. The presence of small bar could also introduce mild distortion, but that is beyond the scope of this study and will be investigated in an upcoming paper which will investigate the non-circular motions induced by the bar as a function of the size and orientation of the bar using numerical simulation.
\section{Mass Models}
\subsection{Mass Models with a Dark Matter Halo}
As mentioned in the introduction, most of the mass in the galaxies appears to be in the form of unseen matter called Dark Matter (DM), whose existence is inferred 
through its gravitational effect on luminous matter such as the flatness of RCs or the gravitational lensing effect. 
The common scenario is that galaxies are embedded in dark matter halos which follow some density distribution profile.
In this paper,
the observationally motivated pseudo-isothermal dark matter halo profile (ISO), characterized by a constant central density core, will be used.
A mass model compares the observed RC to the sum of the contributions of the three mass components, namely the gas disk, the stellar disk 
(and bulge, if present) and the dark halo. It is well known that neutral hydrogen usually has a larger radial extent compared to the luminous part of the galaxy, especially for late--type spirals and dwarf irregular galaxies (at least in the field). Therefore the RC derived using the H{\sc i} gas  traces the mass of the galaxy to larger radii.
The contributions of all the components are required for the mass model. 
The stellar components used in this study are derived from 3.6 micron surface brightness profiles, converted into mass using the mass-to-light ratio (M/L) in that particular band. This M/L is assumed to be constant with radius (cf. section 3.3.2).  
Most of the gas content of the galaxy is in the form of neutral hydrogen, thus the gas contribution is derived from the H{\sc i} maps  corrected for the primordial helium contribution. The dark matter halo component is derived from the pseudo-isothermal density distribution.

The RC is thus given by the quadratic sum of the contribution from each component:
\begin{equation}
V_{rot}^{2} = V_{gas}^{2} + V_{*}^{2} + V_{halo}^{2}
\end{equation}
where $V_{gas}$ is the gas contribution, $ V_{*} $ the stars contribution and $V_{halo}$ the contribution of the dark matter component.


\subsubsection*{The pseudo-Isothermal (ISO) Dark Matter (DM) Halo Model}

For the pseudo-isothermal dark matter halo, the density distribution is given by:
\begin{equation}
\rho_{ISO}(R) =\frac{ \rho_{0}}{1 + (\frac{R}{R_{c}})^{2}}
\end{equation}
while the corresponding RC is given by:
\begin{equation}
V_{ISO}(R) = \sqrt{4\pi G \rho_{0}R_{C}^{2}[1 - \frac{R}{R_{C}}atan(\frac{R}{R_{C}})]}
\end{equation}
where $\rho_{0}$ and  \emph{R$_{c}$} are the central density and the core radius of the halo, respectively.
We can describe the steepness of the inner slope of the mass density profile with a power law $\rho \sim r^\alpha$.
In the case of the ISO halo, where the inner density is an almost constant density core, $\alpha = 0$.
 
\subsection{Mass Models using the MOND formalism}

MOND was proposed by Milgrom as an alternative to dark matter. Therefore, in the MOND formalism only the contributions of the gas and of the stellar component are required to explain the observed RCs.
%

 In the MOND framework, the gravitational acceleration of a test particle is given by :
\begin{equation}
 \mu(x = g/a_{0}) g = g_{N}
\end{equation}
where g is the gravitational acceleration, $\mu$(x) is the MOND interpolating function and g$_{N}$ the Newtonian acceleration.

 \subsubsection{MOND Acceleration ÒConstantÓ a$_{0}$}
 
As an universal constant, a$_{0}$ should be the same for all astrophysical objects. However, observational uncertainties could introduce a large scatter in a$_{0}$ and have to be taken into account. Significant departures of a$_{0}$ from the standard value could be interpreted as being problematic for MOND.
 Milgrom estimated a$_{0}$  using Freeman's law, which stipulates that disk galaxies have typical extrapolated central surface brightness in the B band (Freeman 1970) of 21.65 mag arcsec$^{-2}$. He estimated a$_{0}$  to be $ \sim (0.7 - 3)\times 10^{-8} (M/L)_{*}$ $\rm \ cm \ s^{-2}$. There are other methods which could be used to find a$_{0}$, such as the Tully-Fisher relation. However, the preferred method, used in many studies, is to estimate a$_{0}$ by comparing the computed RCs from mass models to the observed RCs, leaving a$_{0}$ as a free parameter. The implications of  a$_{0}$ in cosmology are explained in \cite{2012LRR....15...10F}.

\subsubsection{MOND Interpolating Functions}
The shape of the predicted MOND RCs depends on the interpolating function. The standard and simple interpolating functions are mostly used in the literature.
The standard $\mu$-function is the original form of the interpolating function proposed by \cite{Milgrom:1983rr} but 
 \cite{2006ApJ...638L...9Z} found that a simplified form of the interpolating function not only provides good fits to the observed RCs  but also the derived mass-to-light ratios are more compatible with those obtained from stellar populations synthesis models.\\
%
%
%
The simple $\mu$-function is given as:

\begin{equation}
\mu(x) = \frac{x}{1 + x}
\end{equation}

%
%
%
%
%
%
%

The MOND RC  using the simple interpolating function is given by

 \begin{equation}
V_{rot}^{2} = \sqrt{V_{*, b}^{2}+ V_{*, d}^{2}  + V_{g}^{2}}*\sqrt{ a_{0}*r + V_{*, b}^{2}+ V_{*, d}^{2}  + V_{g}^{2}}
\end{equation}
where V$_{*, d}$, V$_{*, b}$, V$_{g}$ are the contributions from the stellar disk, the bulge and the gas to the RC.
\subsection{Luminous Component Contribution}
 
\subsubsection{Gas Contribution}

VLA observations using combined B, C and D array configurations data \citep{Walter:2008bs} are used to compute the mass density profile of the  H{\sc i} gas for the THINGS galaxies. It is computed using the GIPSY task {\sc ellint} and the tilted ring kinematical parameters from \cite{de-Blok:2008oq}. The H{\sc i} profile is corrected by a factor of 1.4 to take into account the Helium and other metals. The output from {\sc ellint} is then used in  {\sc rotmod} to calculate the gas contribution in the mass model, assuming an infinitely thin disk. The gas density profiles are taken from \cite{2011MNRAS.410.2217W} for NGC 300 using ATCA data and from \cite{2013AJ....146...48C} for NGC 3109 using KAT7 data,  from \cite{1991AJ....101..447P} for NGC 55 and from \cite{1990AJ....100..641C} for NGC 247, using VLA data. 

\subsubsection{Stellar Contribution}
\label{sec:stellar}

For the mass models, the 3.6 micron surface brightness profiles are used for the stellar contribution. 
3.6 micron probes most of the emission from the old stellar disk population \citep{1997PhDT........13V}. It is also less affected by dust and therefore represents the bulk of the stellar mass. The profiles from \cite{de-Blok:2008oq} are used for the galaxies from the THINGS sample. The profiles were decomposed into two components for the galaxies with a prominent central bulge (see \citealt{de-Blok:2008oq} for more details). The 3.6 micron surface brightness profiles of NGC 55 and NGC 247 are derived in this work. The images were retrieved from the Spitzer Heritage Archive using a 0.6 arcsec/pixel scale. After removing the foreground stars, the images were fitted with concentric ellipses using the {\sc ellipse} task in IRAF. These profiles are corrected for inclination before being converted into mass density. The method from \cite{2008AJ....136.2761O} is adopted to convert the luminosity profiles into mass density profiles for all the galaxies in the sample except for NGC 300 and NGC 3109. \cite{2008AJ....136.2761O} first convert the surface brightness profiles in mag/arcsec$^{2}$ into luminosity density profiles in units of L$_{\odot}$/pc$^{2}$ and then convert to mass density using the following expression:

\begin{equation}
\Sigma[M_{\odot} pc^{-2}] = (M/L)^{3.6}_{*} \times 10^{-0.4 \times (\mu_{3.6} - C^{3.6})}
\end{equation}
where  $(M/L)^{3.6}_{*}$ is the stellar mass-to-light ratio in the 3.6 micron band, $\mu_{3.6}$ the surface brightness profile and C$^{3.6}$ is a constant used for the conversion from mag/arcsec$^{2}$ to L$_{\odot}$/pc$^{2}$.
The details to find C$^{3.6}$ are presented in  \cite{2008AJ....136.2761O} :

\begin{equation}
C^{3.6} = M_{\odot}^{3.6} + 21.56
\end{equation}
where M$_{\odot}^{3.6}$ is the absolute magnitude of the Sun in the 3.6 micron band.
Using the distance modulus formula and the distance to the Sun, \cite{2008AJ....136.2761O} found:
\begin{equation}
M_{\odot}^{3.6} = m_{\odot}^{3.6} + 31.57 = 3.24
\end{equation}
The mass density profile of NGC 300 is taken from  \cite{2011MNRAS.410.2217W}. For NGC 3109, the I-band profile is used instead of the 3.6 microns because it has a larger radial extent.

\subsubsection{Mass to light ratio (M/L)}

The conversion of light into mass through the mass-to-light ratio (M/L) is the principal source of uncertainty in the mass model. Therefore, the determination of this parameter should not be taken lightly or left as a free parameter. In this work, the stellar synthesis models of \cite{2001ApJ...550..212B} are used. As mentioned by \cite{de-Blok:2008oq}, the uncertainties on (M/L) decrease in the infrared band. Therefore, the (M/L) derived in the infrared band will be used when available. 

The method of  \cite{2008AJ....136.2761O} is followed. The (M/L) at 3.6 micron is given by:
\begin{equation}
\log(M/L_{k})=1.46(J - K) -1.38
\end{equation}
and
\begin{equation}
(M/L)_{3.6}=0.92(M/L)_{k} - 0.05
\end{equation}
The J-K colors are those from the 2MASS Large Galaxy Atlas \citep{2003AJ....125..525J}.
An I-band M/L of 0.7 predicted by the stellar population synthesis models is adopted for NGC 3109.

\section{Results}
 The H{\sc i} RCs of the THINGS sample are oversampled with two data points per resolution element.  If we want the reduced $\chi^{2}$  values to be meaningful, the RCs have to be resampled with only one point per beam, so that every point is independent. Therefore,  resampled versions of the THINGS H{\sc i} RCs and of four other RCs from the literature will be used to construct the mass models of the galaxies in our sample. The gas and stellar contributions to the RCs are computed with the GIPSY task {\sc rotmod}. The outputs from {\sc rotmod} are used in {\sc rotmas}, which is the main task for the mass models. {\sc rotmas} uses a non-linear least square method and compares the observed RCs to the calculated RCs derived from the observed mass distribution of the gas and stars \citep{1992ASPC...25..131V}. Inverse squared weighting of the RCs' data points with their uncertainties are used during the fitting procedure. 

\subsection{ISO Dark Matter Halos Fit Results}
Dark matter  mass models are shown in the left panels of Figs. \ref{d1}. The dark matter halo components are shown as dashed magenta lines, the contribution from the stellar disk components as dot-dashed black lines, the stellar bulge components as long-dashed green lines and that from the gas components as dashed red lines. The results are summarized in Table \ref{isonfw}. 

\begin{figure*}
\centering
\begin{minipage}{140mm}
  \begin{tabular}[t]{c}
  	\hspace*{-6.em} \vspace{-2.0em}
        \subfigure{\includegraphics[width=1.23\linewidth]{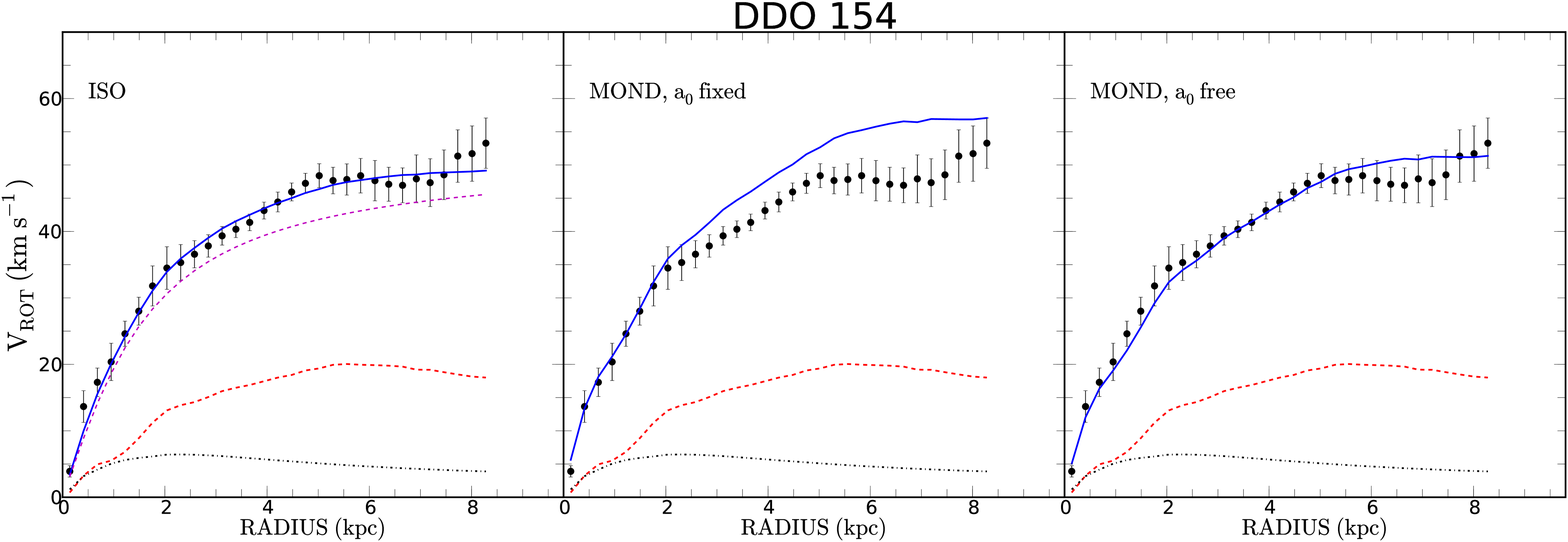}} \\ 
        \hspace*{-6.em} \vspace{-2.0em}
        \subfigure{\includegraphics[width=1.23\linewidth]{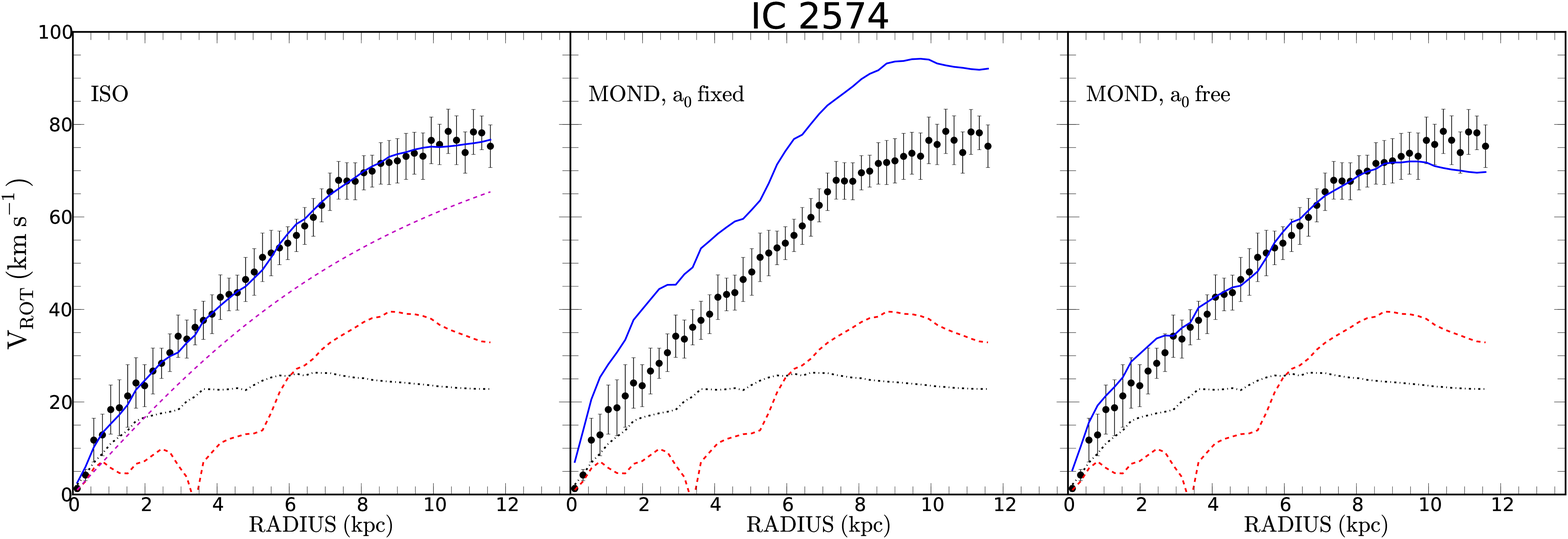}} \\ 
        	\hspace*{-6.em} \vspace{-2.0em}
        \subfigure{\includegraphics[width=1.23\linewidth]{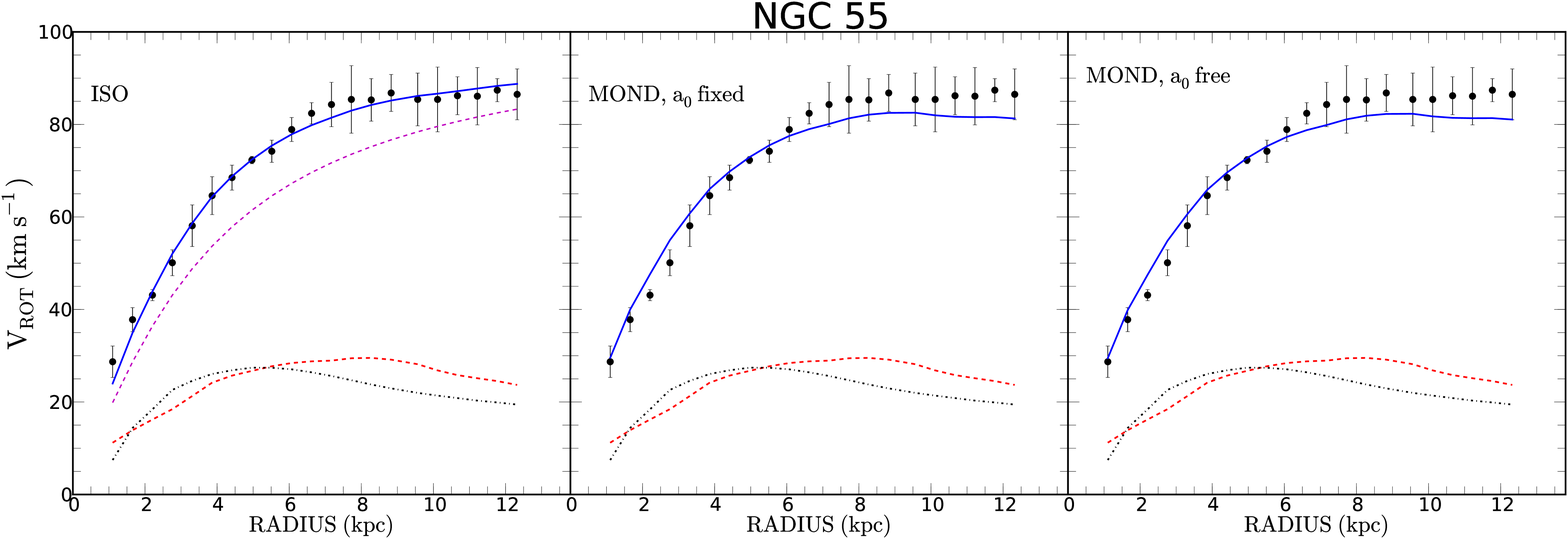}} \\
         \hspace*{-6.em}
        \subfigure{\includegraphics[width=1.23\linewidth]{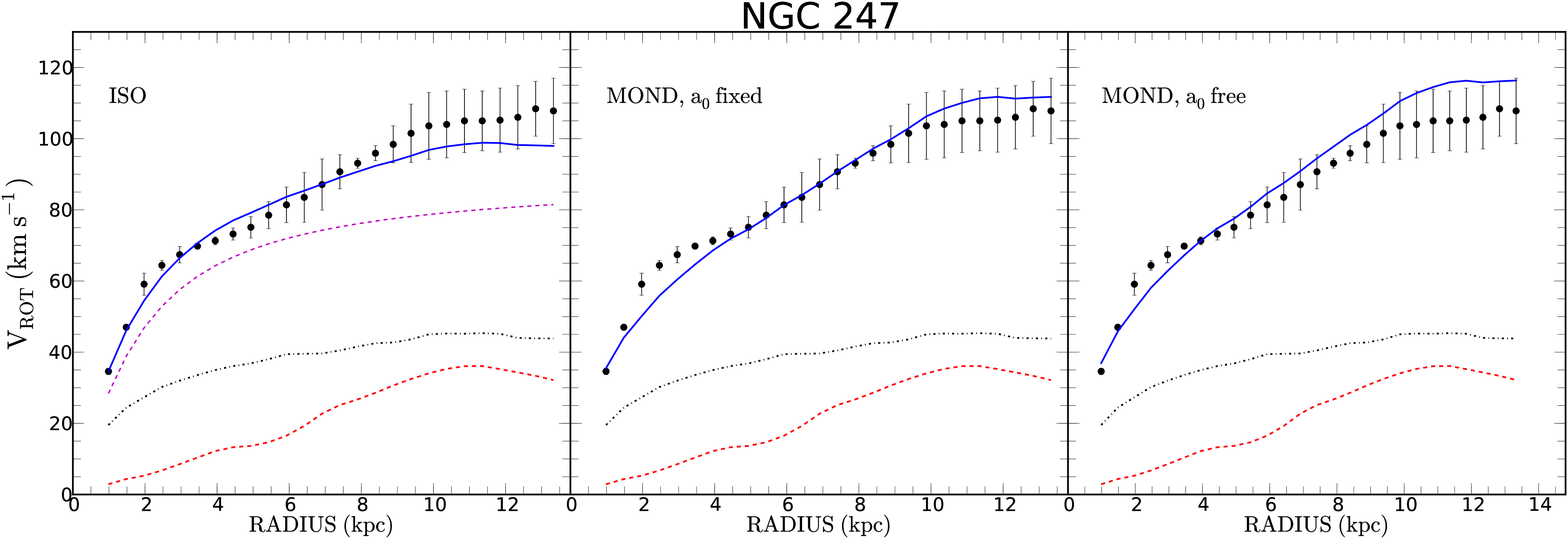}} \\
  \end{tabular}
     \caption[RCs fit results for DDO 154]{Mass model  fit results, left panel: ISO RCs fits (the parameter results are in Table 2), middle panel: MOND RCs fits with a$_{0}$ fixed and right panel: MOND RCs fits with a$_{0}$ free (the MOND parameter results are in Table 3). The red dashed curve is for the \hi\ disk, the dash-dotted black curve is for the stellar disk, dashed green curves for the stellar bulge and the dashed magenta curves for the dark matter component. The bold blues lines are the best--fit models and the black points the observed rotational velocities.}
     \label{d1}
\end{minipage}
\end{figure*}

\begin{figure*}
\centering
\begin{minipage}{150mm}
  \begin{tabular}[t]{c}
  	\hspace*{-6.em} \vspace{-2.0em}
        \subfigure{\includegraphics[width=1.24\linewidth]{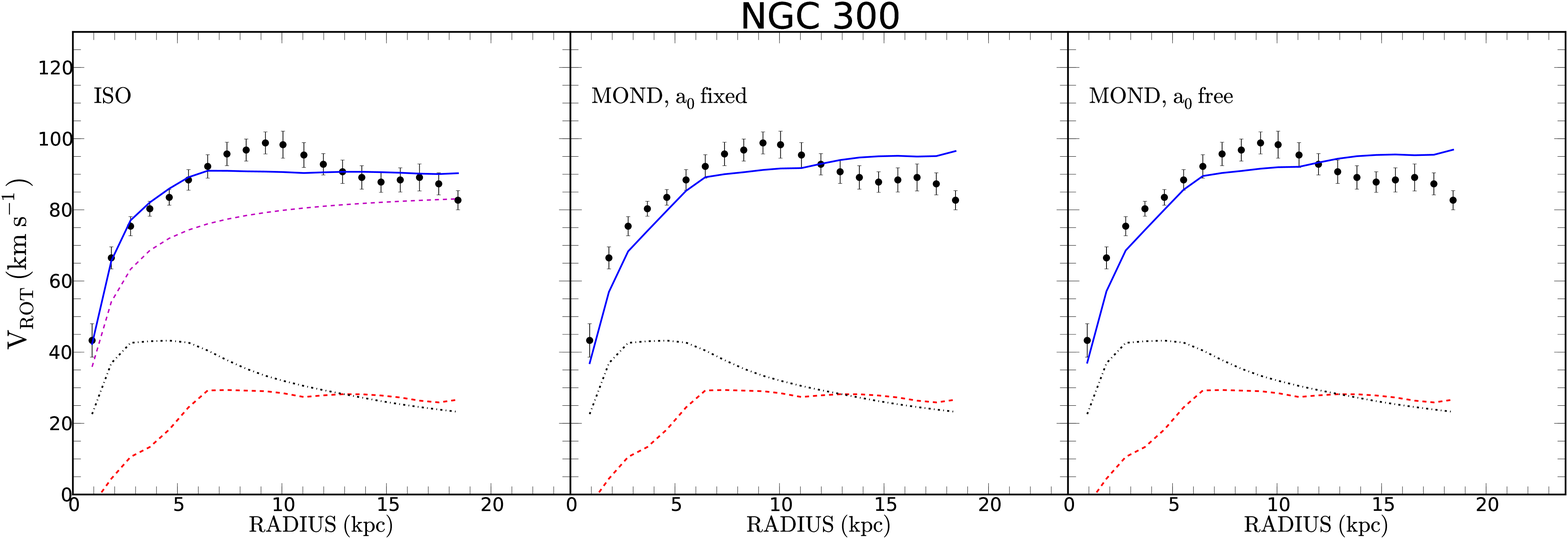}} \\ 
        \hspace*{-6.em} \vspace{-2.0em}
        \subfigure{\includegraphics[width=1.24\linewidth]{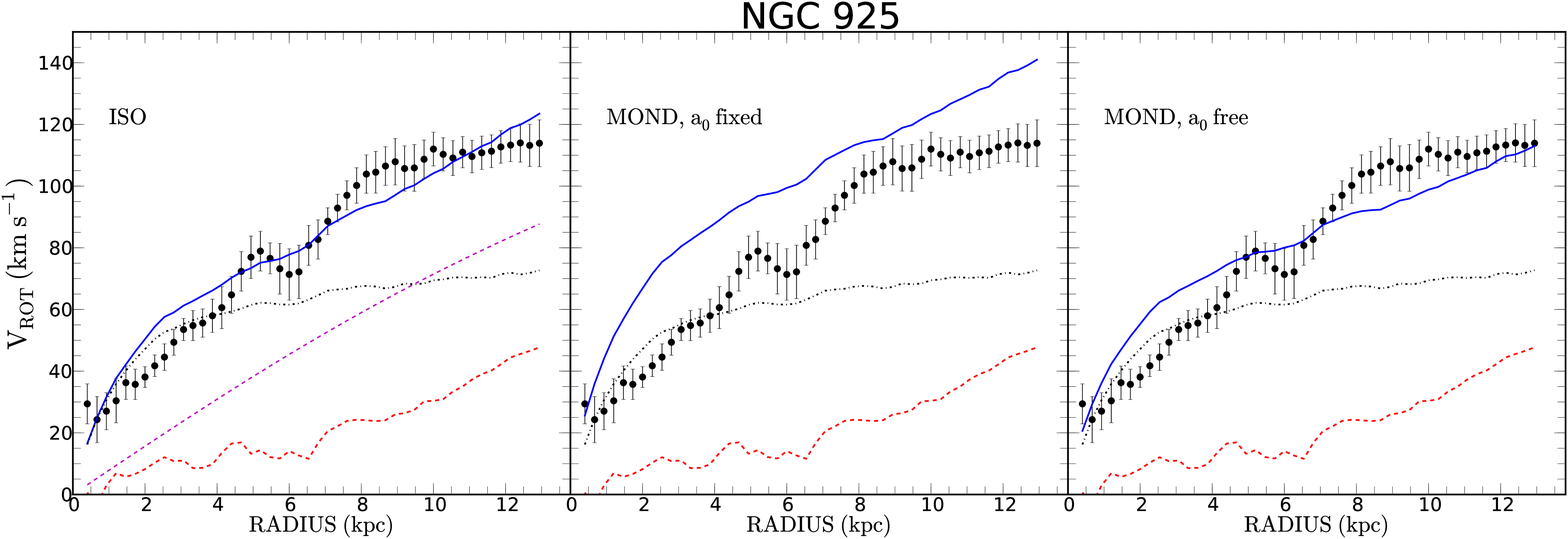}} \\ 
        	\hspace*{-6.em} \vspace{-2.0em}
        \subfigure{\includegraphics[width=1.24\linewidth]{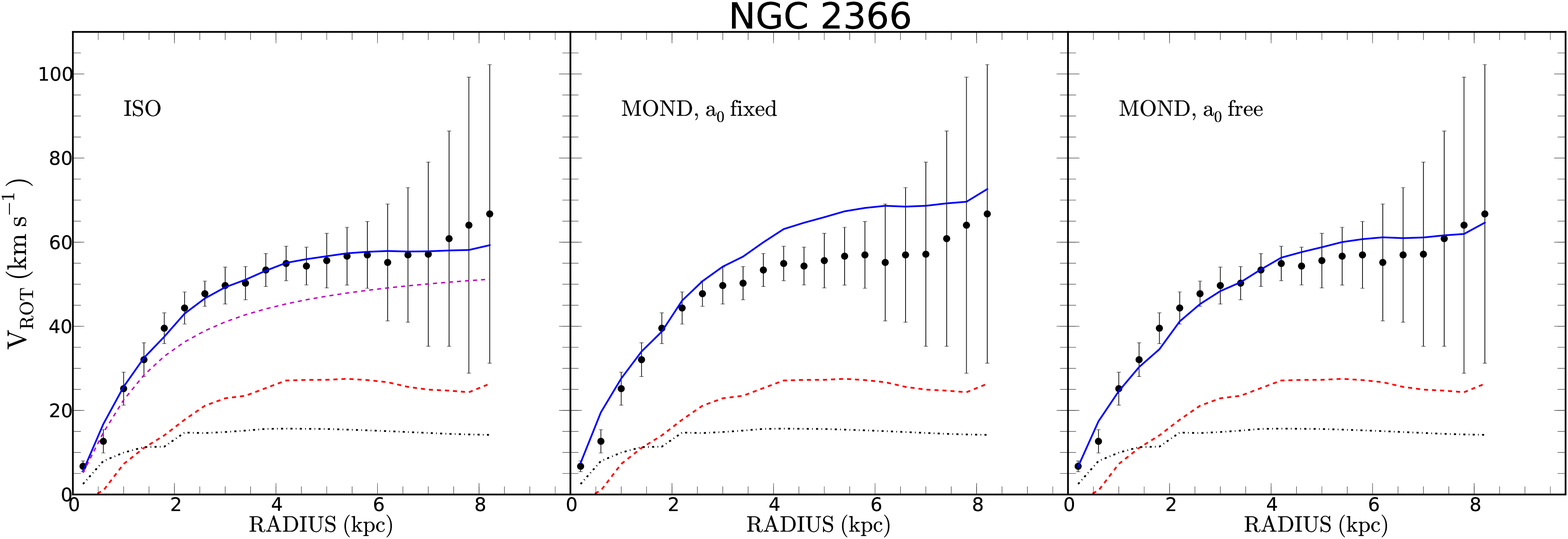}} \\
         \hspace*{-6.em}
        \subfigure{\includegraphics[width=1.24\linewidth]{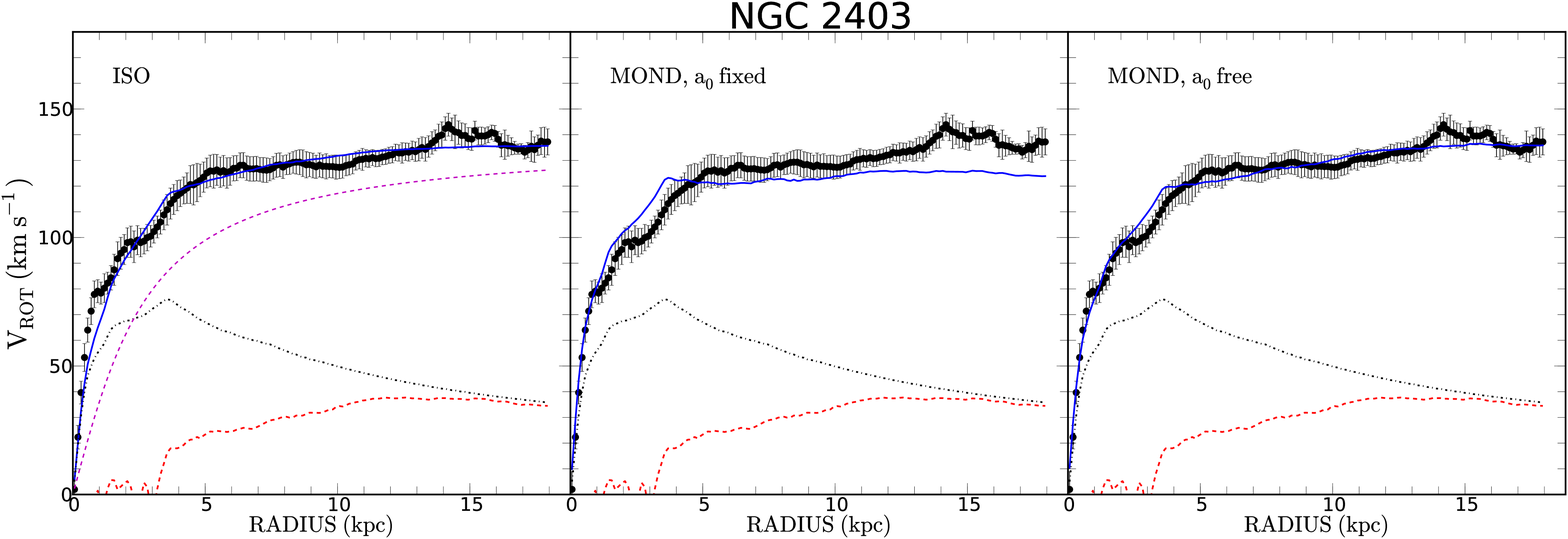}} \\
  \end{tabular}
     \caption*{Continued.}
     \label{d2}
\end{minipage}
\end{figure*}

\begin{figure*}
\centering
\begin{minipage}{150mm}
  \begin{tabular}[t]{c}
  	\hspace*{-6.em} \vspace{-2.0em}
        \subfigure{\includegraphics[width=1.24\linewidth]{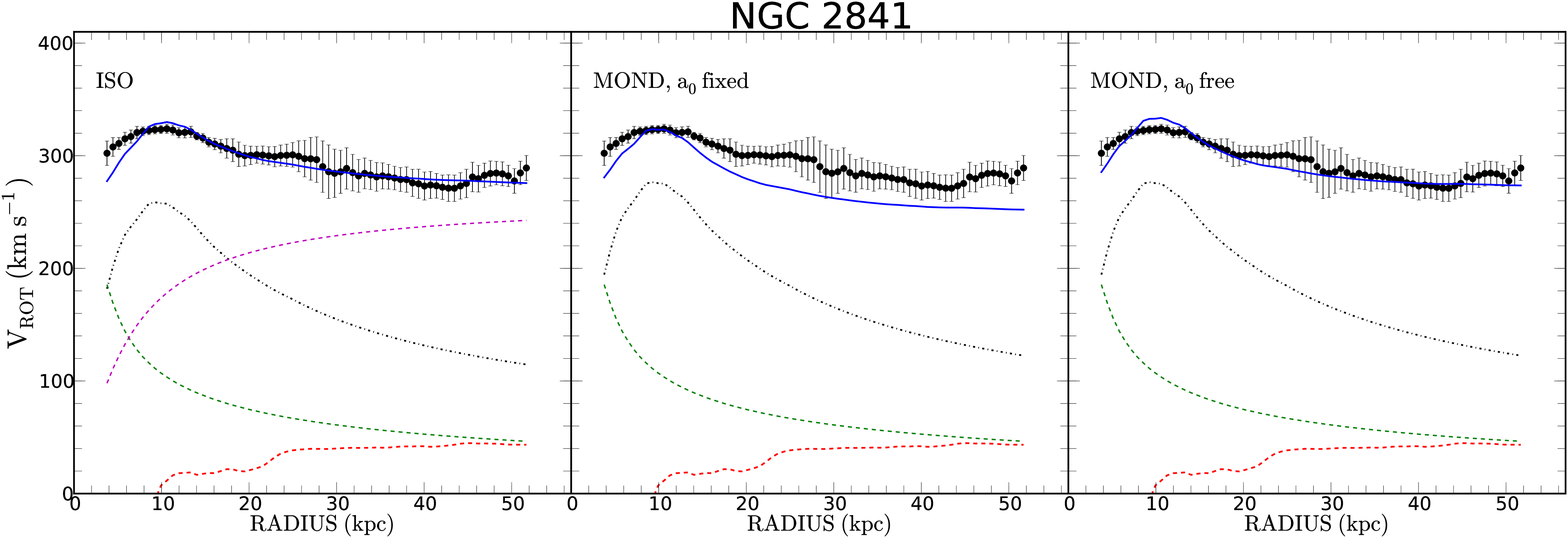}} \\ 
        \hspace*{-6.em} \vspace{-2.0em}
        \subfigure{\includegraphics[width=1.24\linewidth]{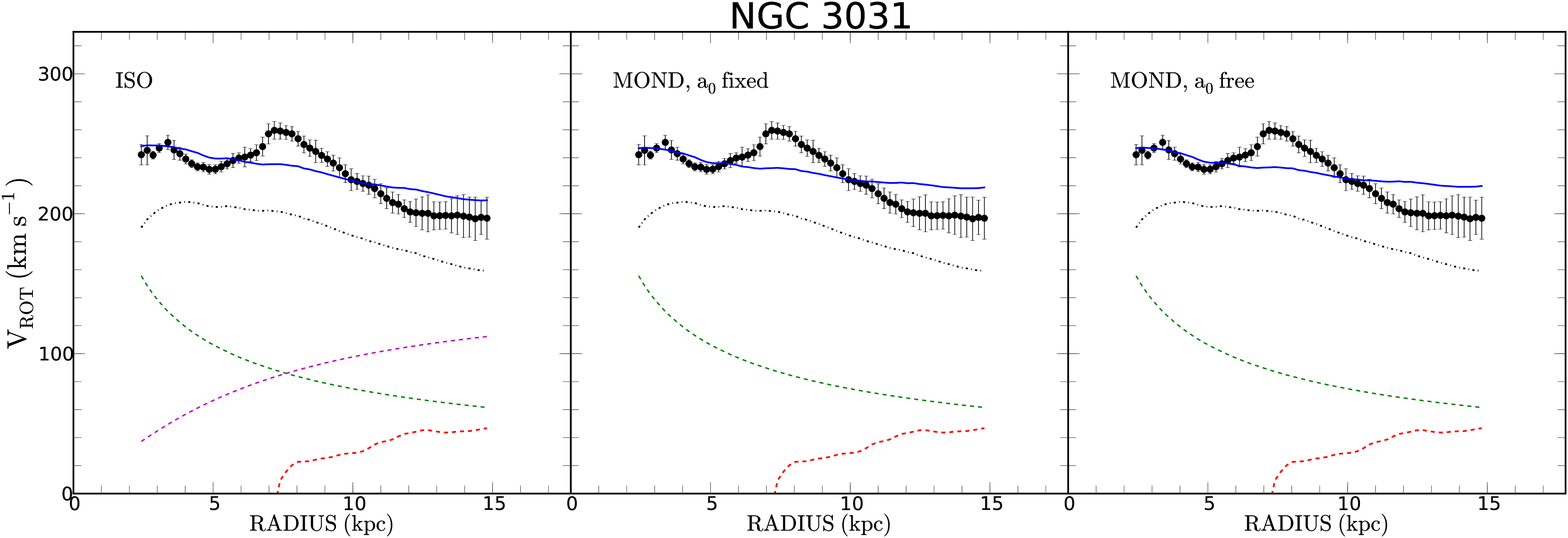}} \\ 
        	\hspace*{-6.em} \vspace{-2.0em}
        \subfigure{\includegraphics[width=1.24\linewidth]{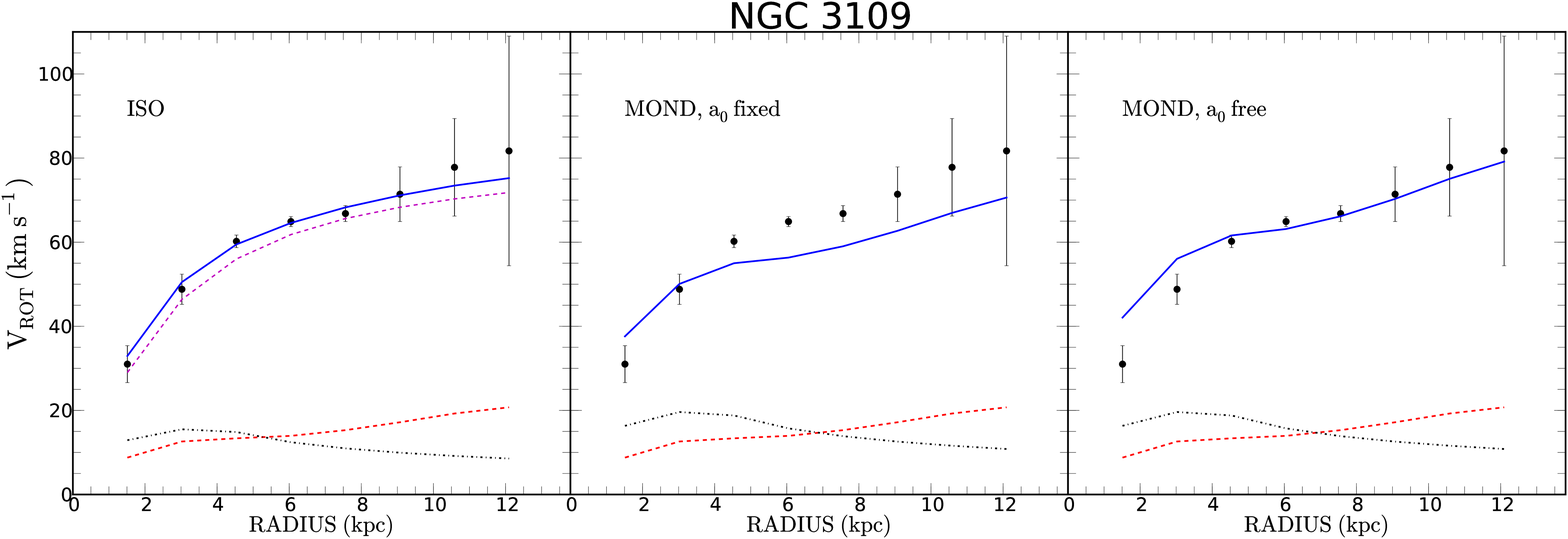}} \\
         \hspace*{-6.em}
        \subfigure{\includegraphics[width=1.24\linewidth]{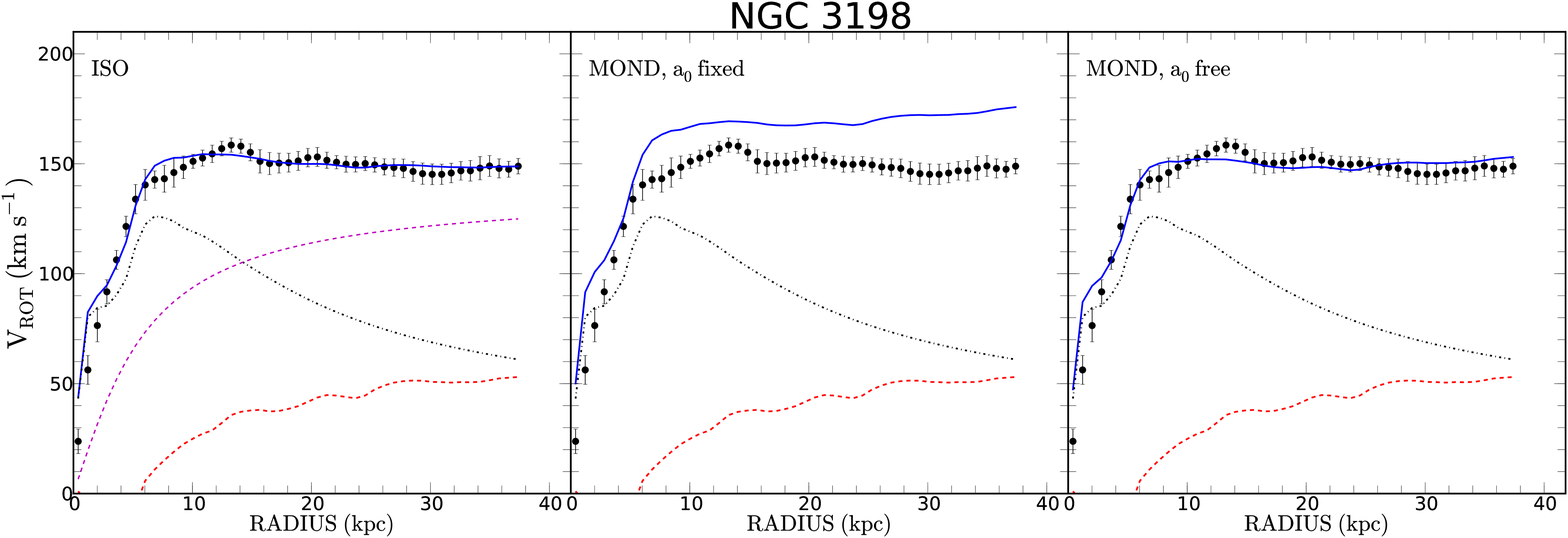}} \\
  \end{tabular}
     \caption*{Continued.}
     \label{d3}
\end{minipage}
\end{figure*}

\begin{figure*}
\centering
\begin{minipage}{150mm}
  \begin{tabular}[t]{c}
  	\hspace*{-6.em} \vspace{-2.0em}
        \subfigure{\includegraphics[width=1.24\linewidth]{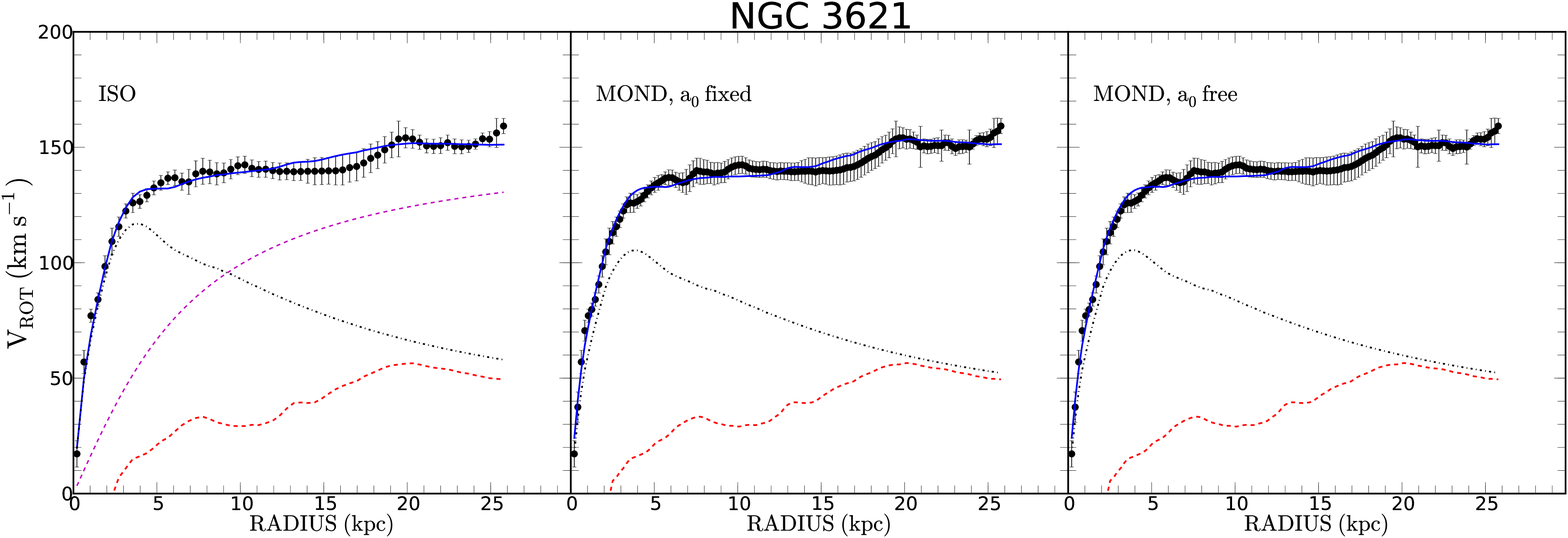}} \\ 
        \hspace*{-6.em} \vspace{-2.0em}
        \subfigure{\includegraphics[width=1.24\linewidth]{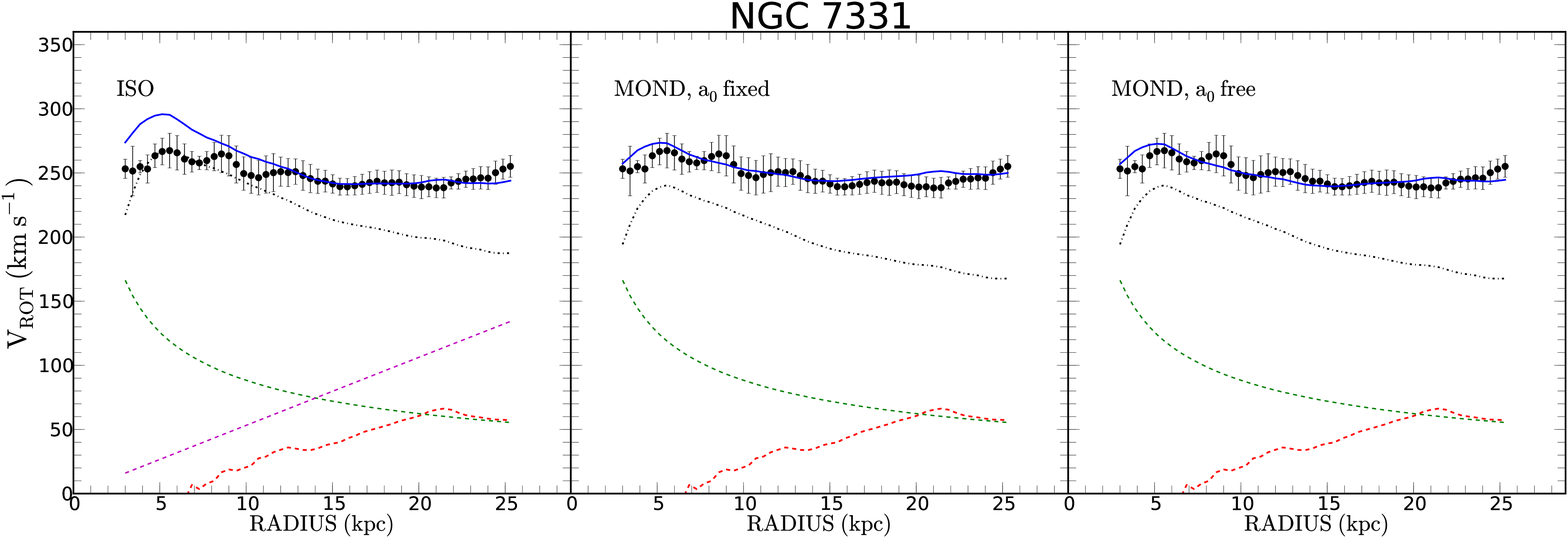}} \\ 
        	\hspace*{-6.em} \vspace{-2.0em}
        \subfigure{\includegraphics[width=1.24\linewidth]{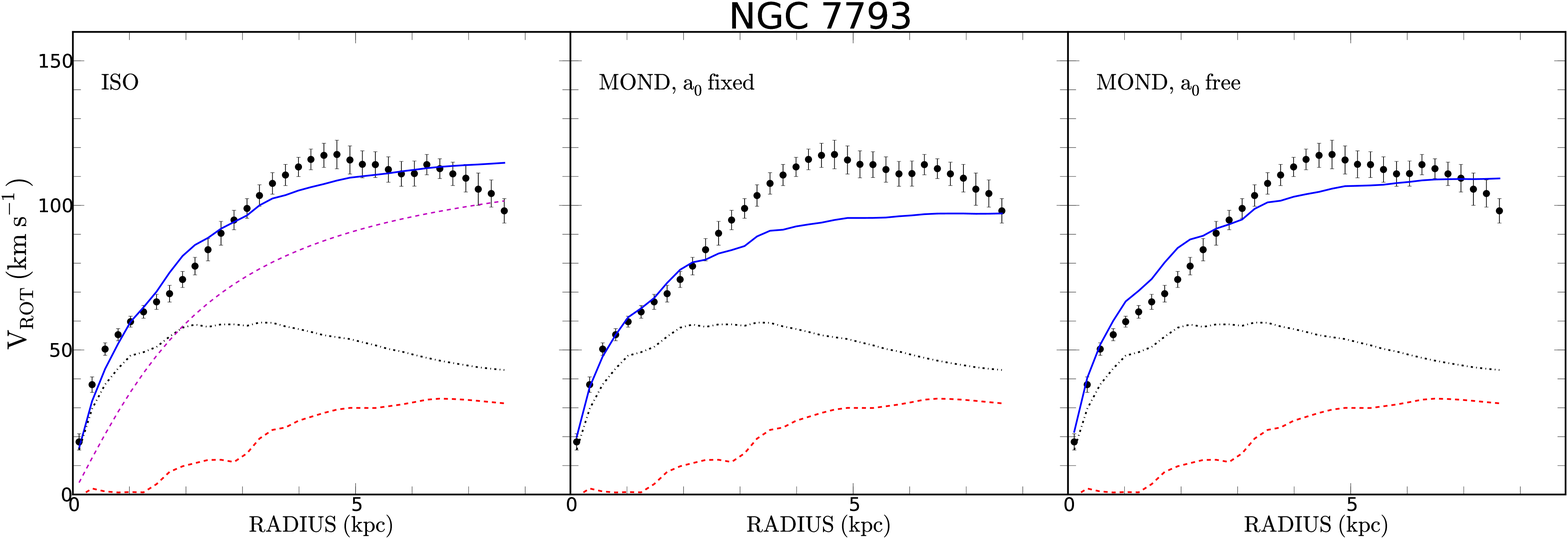}} \\
         \hspace*{-6.em}
      \end{tabular}
     \caption*{Continued.}
     \label{d4}
\end{minipage}
\end{figure*}

\subsection{MOND Results}
\subsubsection{MOND fits with distance fixed}
The MOND models were performed with the same (M/L)s as the ones used for the DM models.
MOND fits  are shown in the middle panels of  Figs. \ref{d1} for a$_{0}$ fixed at its standard value and in the right panels for a$_{0}$ free. The observed RCs are shown as black points with error-bars and the MOND RCs calculated from the observed mass density distribution of the stars and gas in continuous blue lines. The stellar disk and bulge contributions are shown as black dot-dashed and long-dashed lines respectively and the gas contributions as red dashed lines.  
Results for the a$_{0}$ fixed and a$_{0}$ free fits are also shown in Table \ref{isonfw}.
An average value for a$_{0}$ of (1.13 $\pm$ 0.50)$\times$10$^{-8}$ cm s$^{-2}$ 
is found using the simple interpolating function. As for \citet{Bottema:2002kl}, this is smaller than the standard value of  \cite{1991MNRAS.249..523B}. 
\subsubsection{MOND fits with distance let free to vary within the uncertainties.}
The distance was allowed to vary within the uncertainties in the case of galaxies in which MOND produces poor fits to their RCs. The results are shown in Table \ref{mond}. 

\begin{table*}
\centering
\scriptsize
\begin{minipage}{180mm}
\caption{Results for the ISO dark matter halo models with fixed M/L (Diet-Salpeter IMF) and for the MOND models using $a_{0}$ = 1.21x10$^{-8}$ cm s$^{-2}$ and $a_{0}$ as a free parameter (simple interpolating function).}
\begin{center}
\begin{tabular}{c c  c  c   c    c    c    c    c    c c  }
\hline\hline 
& & &&&ISO&&&&MOND&\\
\cline{5-7} \cline{9-11}
Name& (M/L)$_{3.6,disk}$&(M/L)$_{3.6,bulge}$ && R$_{C}$ &$\rho_{0}$ &$\chi^{2}_{r}$&&$\chi^{2}_{r}$ (a$_{0}$ fixed)& a$_{0}$ &$\chi^{2}_{r}$ (a$_{0}$ free)\\\\

 & & & & kpc& 10$^{-3}$ M$_{\odot}$ pc$^{-3}$&&&&10$^{-8}$ cm s$^{-2}$\\\\
  1&2&3&  & 4&5&6&&7&8&9\\
\hline \\
DD0 154 &0.32[dB08]& & & 1.34$\pm$0.07 &27.38$\pm$2.33&0.42&&6.32&0.68 $\pm$0.02&0.56\\\\

IC 2574 &0.44[dB08]&& & 7.36$\pm$0.21 & 4.02$\pm$0.22&0.22&&19.28&0.36 $\pm$ 0.02&1.87 \\\\

NGC 0055 &0.44[Tw]& & & 3.17$\pm$0.12 &19.19$\pm$0.88& 0.36&&&1.15 $\pm$ 0.05 & 1.78\\\\

NGC 0247 &0.36[Tw]& & & 1.63 $\pm$0.08 &55.45 $\pm$3.88 &1.59&&6.06   &1.43 $\pm$ 0.05 & 3.31 \\\\

NGC 0300 &0.27[W11]& & & 1.08$\pm$0.15 &117.93 $\pm$ 29.63 & 1.78&&5.43&1.18 $\pm$ 0.08 & 4.55\\\\

NGC 0925 &0.65[dB08]&& & 16.63$\pm$10.16&3.40 $\pm$ 0.74 & 2.09&&22.09&0.34 $\pm$ 0.04 & 4.59\\\\

NGC 2366 &0.33[dB08]& && 1.28$\pm$0.11 &37.47 $\pm$ 4.25 & 0.20&&2.11&0.71 $\pm$ 0.04 & 0.39\\\\

NGC 2403 &0.74[dB08]&&&  4.53$\pm$0.15&20.97 $\pm$ 1.02& 0.63&&4.74& 1.51$\pm$ 0.03 & 2.72\\\\

NGC 2841 &0.74[dB08]&0.84& &  5.08$\pm$0.23&49.06$\pm$3.61& 0.82&&4.69&1.72 $\pm$ 0.03 & 0.96\\\\
NGC 3031  &0.80[dB08]&1.00& &5.34$\pm$1.97&14.55$\pm$5.87&3.97&&4.77&1.24$\pm$ 0.09 &4.52 \\\\

NGC 3109&0.70$^{a}$[Tw]& & &  2.22$\pm$0.20&25.71$\pm$3.21&0.25&&21.24&1.91 $\pm$0.14&1.94\\\\

NGC 3198 &0.80[dB08]&&  & 4.85$\pm$0.42&15.01 $\pm$ 2.15& 1.17&&24.26&0.67 $\pm$ 0.02 & 1.84\\\\

NGC 3621 &0.59[dB08]& &&  5.56$\pm$0.23& 14.31$\pm$0.16& 0.70&&1.56& 0.98 $\pm$ 0.02& 1.52\\\\

NGC 7331 & 0.83[dB08]&1.00& & 17.38$\pm$2.75 & 4.75$\pm$0.60&0.45&&0.68& 1.08 $\pm$ 0.03& 0.42\\\\

NGC 7793 &0.31[dB08]&& & 1.90$\pm$0.20& 77.95$\pm$11.13&3.06&&12.58& 2.01 $\pm$ 0.13& 4.63\\\\
   &&&  &  &&   \textless 1.18\textgreater&&\textless 9.20 \textgreater&\textless 1.13 $\pm$ 0.50 \textgreater&\textless 2.37 \textgreater\\\\
\hline

\end{tabular}
\end{center}
\addtocounter{footnote}{-2}
$^{a}$\footnotesize{the I-band surface brightness profile was adopted for NGC 3109 because it has larger radial extent than the 3.6 microns surface brightness profile.}\\
 \footnotesize{col. 1: Galaxy name}; \quad
   \footnotesize{col. 2 \& 3:  mass-to-light ratio [reference: dB08: \cite{de-Blok:2008oq}
; Tw: this work; W11: Westmeier et al. (2011)] }; \quad
\footnotesize{col. 4: ISO halo core radius}; \quad
 \footnotesize{col. 5: ISO halo central density}; \quad
 \footnotesize{col. 6: ISO reduced chi-squared.}\quad
 \footnotesize{col. 7 \& 9:  reduced chi-squared for a$_{0}$ fixed and free }; \quad
  \footnotesize{col. 8:  MOND acceleration parameter  }\\
\label{isonfw}
\end{minipage}
\end{table*}

\begin{table*}
\centering
 \begin{minipage}{140mm}
\caption{Effect of varying the adopted distance within the uncertainties (M/L fixed (Diet-Salpeter  IMF, $a_{0}$ = 1.21x10$^{-8}$ cm s$^{-2}$ and using the simple interpolating function).}
\begin{center}
\begin{tabular}{ c      c     c  c    c   }
\hline\hline 
Name& (M/L)$_{3.6, disk}$  & (M/L)$_{3.6, bulge}$&Distance&$\chi^{2}_{r}$ \\
 &  &&Mpc &  \\
 1 & 2 & 3 &4&5 \\
\hline \\
DD0 154 & 0.32&&4.30&6.32\\\\
 & &&3.76&3.21\\\\
  & &&3.23&1.21\\\\
IC 2574 & 0.44& &4.02&19.28 \\\\
 & & &3.61&14.84  \\\\

NGC 0247 & 0.36 &&3.41& 6.06 \\\\
 &  &&3.58& 4.96  \\\\
NGC 0300 & 0.27 &&1.99& 5.43 \\\\
 & &&1.96& 4.55 \\\\
NGC 0925 & 0.65  &&9.16& 22.09 \\\\
 &   &&8.53& 18.48  \\\\
NGC 2403 &0.74  && 3.22&4.74\\\\
 &  && 3.36&4.06 \\\\
  &  && 3.47 [G11]&3.60 \\\\
   &  && 3.76&2.88 \\\\
NGC 2841 & 0.74  &0.84& 14.10&4.69\\\\
 &   && 15.60&2.13 \\\\

NGC 3198 & 0.80  & &13.80&24.26 \\\\
 &  & &12.95&17.43  \\\\
 &  & &12.30&14.36  \\\\

NGC 7793 &0.31  & &3.43&12.58\\\\
 &  & &3.53&10.88\\\\
 &  & &3.91[G11]&8.55 \\\\
 &  & &4.30&6.89 \\\\
\hline
\end{tabular}
\end{center}
\addtocounter{footnote}{-2}

 \footnotesize{col. 1: Galaxy name}; \quad
 \footnotesize{col. 2 \& 3: mass-to-light ratio}; \quad
  \footnotesize{col. 4: adopted distances, G11: distance used by Gentile et al. (2011) (see text for more explanation)}; \quad
 \footnotesize{col. 5 :  reduced chi-squared }; \quad
\label{mond}
\end{minipage}
\end{table*}

\section{Discussion}

The average reduced chi-square for the ISO halo ( \textless$\chi^{2}_{r}$\textgreater = 1.18 ) is lower than that obtained from MOND with fixed a$_{0}$ (\textless$\chi^{2}_{r}$\textgreater =  9.20) using the distance listed in Table 1 and even when a$_{0}$ is allowed to vary (\textless$\chi^{2}_{r}$\textgreater =  2.37). 
Discrepancies between the RCs predicted by MOND and the observed RCs are seen for most of the galaxies in the sample and particularly for DDO 154, IC 2574, NGC 925, NGC 2841, NGC 3109, NGC 3198 and NGC 7793, in which MOND overestimates or underestimates the rotation velocities. 
The difference between the MOND fits and the observed RCs is smaller for the following galaxies: NGC 0055, NGC 2366, NGC 3621 and NGC 7331 when a$_{0}$ is fixed to its canonic value. Notes on the individual galaxies are given in the appendix, where the results from this study are compared to those in the literature. 

\subsection{Effect of varying the adopted distances within the uncertainties}
The results are summarized in table \ref{mond}. The adopted distances  are different with \cite{Gentile:2011th} for the following galaxies: DDO 154 (the error-bars are not the same), NGC 2403, NGC 3198 (the error-bars are not the same) and NGC 7793. MOND prefers lower distances for most of the galaxies except for NGC 247,  NGC 2403, NGC 2841 and NGC 7793. It is worth noticing that a much lower distance is needed for some galaxy such as NGC 3198 if the M/L is fixed.
The quality of the fits improves when the MOND acceleration constant a$_{0}$ is taken as a free parameter for all the galaxies.

\subsection{Comparison with previous work}

There are a wealth of studies on MOND in the literature, but this section will focus on a comparison between this work and that of \cite{Gentile:2011th} because, not only \cite{Gentile:2011th} is the most recent study, but both samples have a large overlap. The main difference between this work and \cite{Gentile:2011th} is that in this work, MOND produces acceptable fits for $\sim$60\% of the galaxies in the sample, while \cite{Gentile:2011th} found that MOND produces excellent fits for all the galaxies except for three (DDO 154, NGC 2841 and NGC 3198), meaning for $\sim$75\% of their sample. The similarity between this paper and \cite{Gentile:2011th} is that seven galaxies are common in both studies. For those seven galaxies, five galaxies produce acceptable MOND fits when a$_{0}$ was left free to vary.
The differences stem from two main reasons:
\begin{itemize}
 \item First, \cite{Gentile:2011th} allowed the mass-to-light ratio to vary and the distances were constrained within the errors, which gave smaller chi-squared values. In this work, the mass-to-light ratios are fixed using those found using population synthesis models.
 \item Secondly, the distance used in this work and \cite{Gentile:2011th} are not the same, for example the distance adopted in this work for NGC 2403 is (3.22 $\pm$ 0.14) Mpc, as listed in Table 1 while their distance is (3.47 $\pm$ 0.29) Mpc. Our distances are mainly cepheid-based.
\end{itemize}

However, this work would reach the same conclusions as \cite{Gentile:2011th} if the mass-to-light ratio was also a free parameter and the distances only constrained. It is well known that the more you have free parameters, the easier it is to get good fits. This is why \cite{Gentile:2011th} get 75 \% (9/12) good fits with their distance constrained option and free mass-to-light ratios, while this falls to 60 \% (9/15) when the distance and mass-to-light ratio are fixed. The difference between the two studies is probably not significant given the small sample sizes in both studies.

\subsection{Correlation Between the MOND Acceleration Constant a$_{0}$ and other Galaxy Parameters}

\begin{figure}
\centering
   \begin{center}
  \begin{tabular}[t]{c}
      \subfigure{\includegraphics[scale = 0.34]{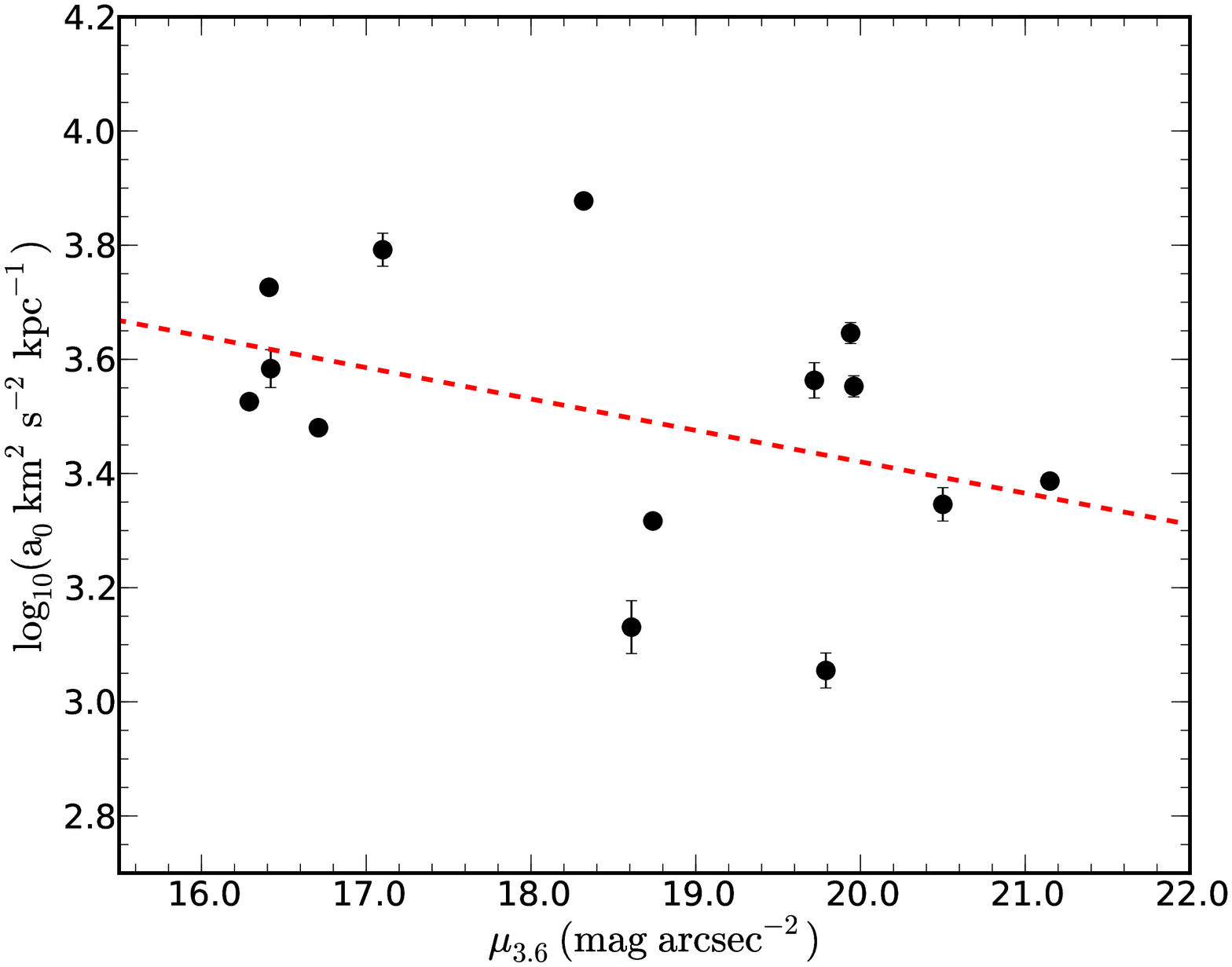}}
  \end{tabular}
     \caption[RC  fit resuts for DDO 154]{MOND parameter as a function of the corrected central surface brightness of the stellar disk in the 3.6 micron band.}
     \label{corrsb}
  \end{center}
\end{figure}

Any systematic trend of a$_{0}$ with some galaxy parameter could be a problem for MOND since it is supposed to be a universal constant. 
Here, we look for a correlation between the corrected central surface brightness of the stellar disk with the MOND parameter a$_{0}$. As shown in Fig. \ref{corrsb}, galaxies with higher central surface brightness require higher values of a$_{0}$ and galaxies with lower central surface brightness lower a$_{0}$ values. This has also been seen in the R-band for LSB galaxies by \cite{Swaters:2010hc}. 
\cite{Gentile:2011th} did the same analysis using the 3.6 micron band for twelve (12) galaxies from the THINGS sample and did not find any correlation. However, the bulge central surface brightnesses were used for their study instead of the disk values. Five galaxies in their sample (see their figure 2) have central surface brightnesses brighter than 13 mag/arcsec$^{2}$ which are probably due to the small but bright central bulges. The surface brightness profile increases sharply within a small radius and lead to a very high estimated central surface brightness, which is the reason why the extrapolated disk central surface brightness have to be used.
The following relation is found in this work which is shown in Fig. \ref{corrsb} as a dashed line. 
\begin{equation}
\log(a_{0}) = (-0.055 \pm 0.037)\times \mu_{3.6} + (4.520 \pm 0.687)
\end{equation}

This surely challenges the universality of MOND, since a$_0$ should be the same for every galaxy.

 \subsection{Dark Matter Halo Scaling Laws}

\begin{figure}
\centering
   \begin{center}
  \begin{tabular}[t]{c}
  \hspace{-2.0em}
      \subfigure{\includegraphics[scale = 0.32]{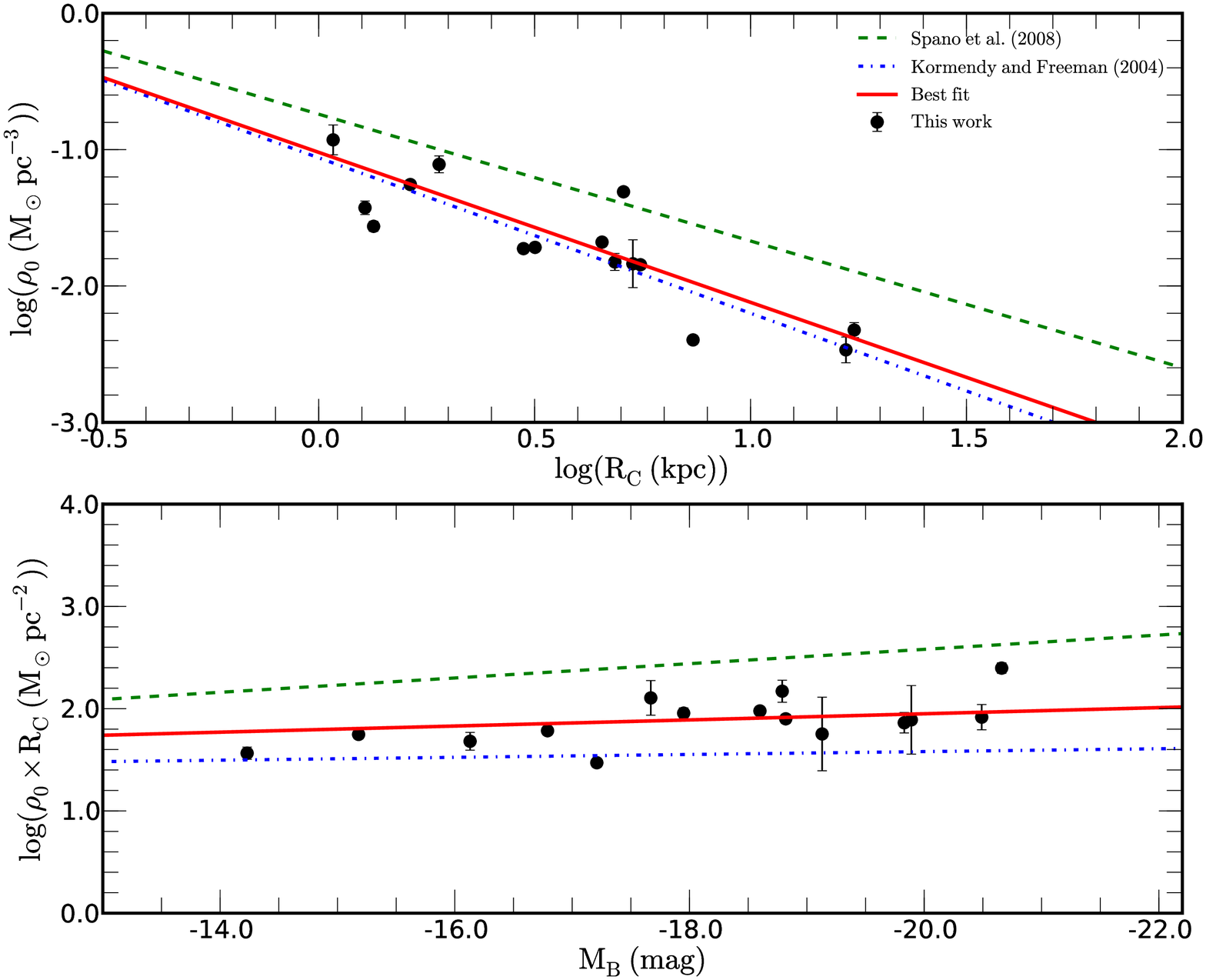}}
  \end{tabular}
     \caption[RC  fit resuts for DDO 154]{Top panel: Core radius as a function of central density  for the ISO halo. The bold red lines show the best fit result found in this study, the dashed green lines are the correlation found by \cite{2008MNRAS.383..297S}, the  dot-dashed blue lines are the correlation found by \cite{2004IAUS..220..377K} and the filled black circle are the results from this work . Bottom panel: halo surface density as a function of B-band absolute magnitude.}
     \label{mbr}
  \end{center}
\end{figure}

The most common hypothesis to explain the flatness of galaxies' RCs is to postulate the existence of a dark matter halo. The halo is characterized by a theoretical density profile. It has been known that the observational motivated ISO halo provides a better description of the observed RCs as compared to the cosmological motivated NFW halo (see e.g. \citealt{2001AJ....122.2396D}). The success of the ISO halo for fitting galaxy RCs has been said to be due to the number of parameters involved in the fitting procedure. These parameters  are the halo core radius \emph{R$_{C}$} and the halo central density $\rho_{0}$. A correlation between these two parameters have been investigated in the literature \citep{2004IAUS..220..377K, 2004AJ....128.2724B, 2008MNRAS.383..297S}. \cite{2004IAUS..220..377K} found the following relation:

\begin{equation}
\label{ }
\log \ \rho_{0} = -1.04 \times log \ R_{C} - 1.02
\end{equation}
while \cite{2008MNRAS.383..297S} found:

\begin{equation}
\label{ }
\log \ \rho_{0} = -0.93 \times log \ R_{C} - 0.74
\end{equation}

\begin{figure}
 \begin{center}
  \begin{tabular}[t]{c}
  \hspace{-2.4em}
      \subfigure

{\includegraphics[scale = 0.32]{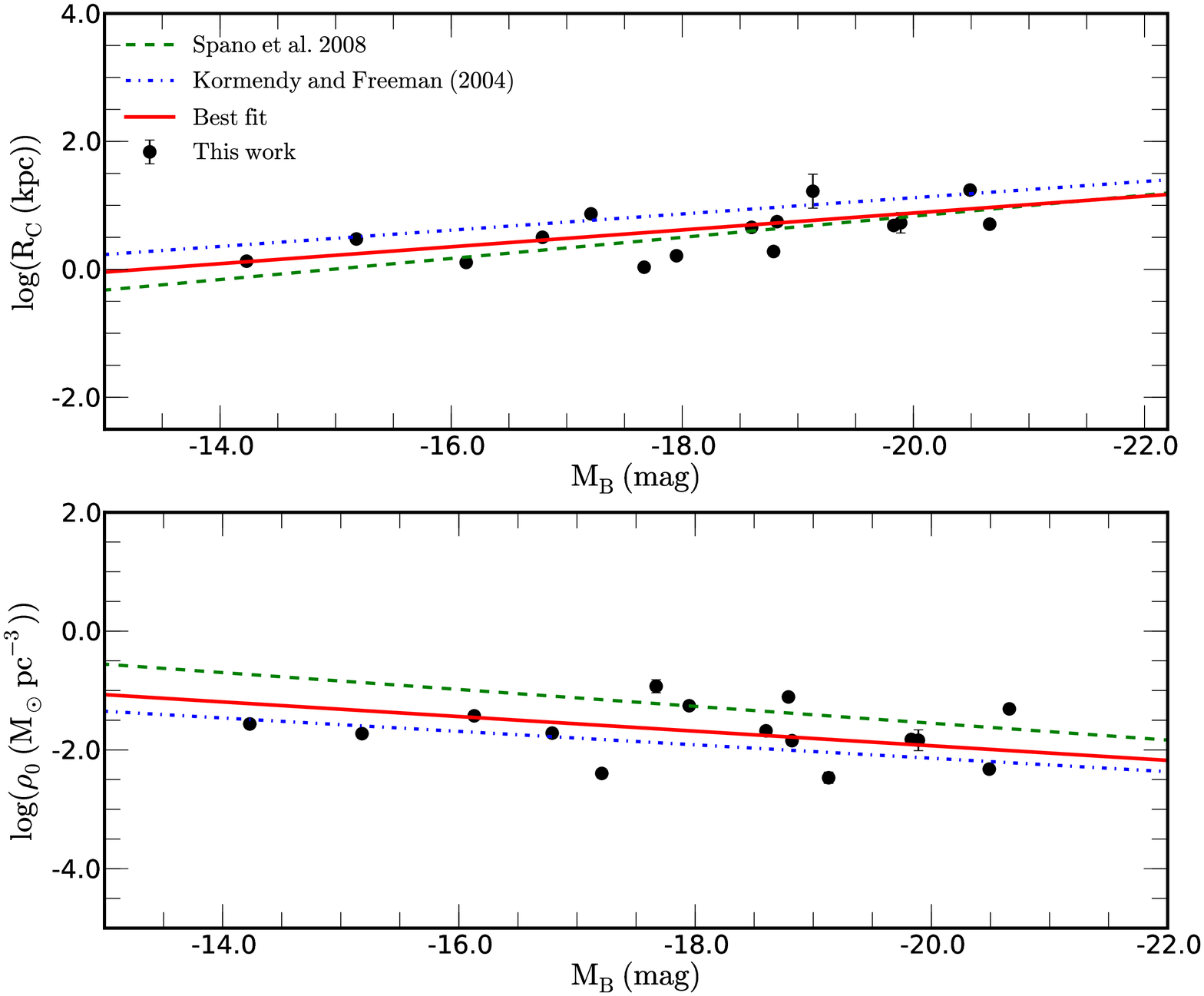}}\\

 \end{tabular}
   \caption[Core radius and central density (log) of the halo as a function of absolute magnitude (ISO model) ]{Top panel: Core radius of the ISO halo as a function of absolute magnitude. Bottom panel: central density as a function of absolute magnitude. The bold red  lines show the correlation found in this study, the long-dashed green lines are the correlation found by \cite{2008MNRAS.383..297S}, the dot-dashed blue lines are the correlation found by \cite{2004IAUS..220..377K} and the filled black circle are the parameter results from this work . }
    \label{mbrc}
  \end{center}
\end{figure}

A similar analysis is undertaken for our sample.  A plot of the core radius as a function of the central densities is shown on the top panel of Fig. \ref{mbr}. We found that our result is consistent with those presented in the literature. This confirms the existence of a scaling relation for the central core of the dark matter halos.
 The following relationship was found using a simple least square method: 
 
\begin{equation}
\label{ res}
\log \ \rho_{0} = (-1.10 \pm 0.13) \times log \ R_{C} - (1.05 \pm 0.25)  
\end{equation}
The core radius is plotted as a function of absolute magnitude in Figure \ref{mbrc}. A clear correlation is seen between these two parameters: low-luminosity galaxies correspond to a small core radius. 

As shown in figure \ref{mbrc} our least square fit results are :
\begin{equation}
\label{ res}
\log \ \rho_{0} = (0.121 \pm 0.021) \times M_{B} + (1.531 \pm 0.120)
\end{equation}
\begin{equation}
\label{ res1}
\log \ r_{c} = (-0.133 \pm 0.044) \times M_{B} - (1.76 \pm 0.797)
\end{equation}
The least square result found by \cite{2008MNRAS.383..297S} were:

\begin{equation}
\label{ res}
\log \ \rho_{0} = 0.142 \times M_{B} + 1.29
\end{equation}
\begin{equation}
\label{ res1}
\log \  r_{c} = -0.167 \times M_{B} - 2.47
\end{equation}
and the results by Kormendy \& Freeman (2004) were:

\begin{equation}
\label{ res}
\log \ \rho_{0} = 0.113\times M_{B} + 0.12
\end{equation}
\begin{equation}
\label{ res1}
\log \ r_{c} = -0.127 \times M_{B} - 1.42
\end{equation}

The correlation between the central density and the absolute magnitude was also investigated which is shown in bottom panel of Fig. \ref{mbr}, this is in good agreement with \cite{2004IAUS..220..377K}.

The halo surface density is given by the product of the core radius and the central density of the halo. \cite{2004IAUS..220..377K} found that the surface density of the halo is nearly constant as a function of absolute magnitude. This result was confirmed by \cite{2008MNRAS.383..297S} which is shown as a green dashed line in figure \ref{mbr}. We found a correlation which is in good agreement with those found in the literature.

\noindent \cite{2004IAUS..220..377K}:
\begin{equation}
\label{ res}
 \rho_{0} \ R_{C} \sim 100 \ M_{\odot} \ pc^{-2}
\end{equation}
Spano et al. (2008):
\begin{equation}
\label{ res}
 \rho_{0} \ R_{C} \sim 150 \ M_{\odot} \ pc^{-2}
\end{equation}
This work:
\begin{equation}
\label{ res}
 \rho_{0} \ R_{C} \sim 120 \ M_{\odot} \ pc^{-2}
\end{equation}

These results confirm the scaling laws for dark matter halos, which is important for our understanding of the relation between the dark and luminous matter and the characteristics of the dark matter itself. It clearly confirms that low luminosity galaxies have smaller core radii and higher central densities. Most importantly, it mainly implies that dark matter ISO halo could be characterized by only one parameter since the core radius and the central density of the halo are correlated. With the (M/L) of the disk now fixed by population synthesis models, DM models are thus left with only one free parameter the DM surface density. 
\section{Conclusions}
We have presented mass models of fifteen (15) dwarf and spiral galaxies selected from the literature. Their observed RCs were confronted with MOND and the observationally motivated ISO dark matter halo model. The galaxies in the sample were selected to be homogenous in terms of their measured distances, the sampling of their RCs and the stellar M/L of the stellar contributions. The selected galaxies in the sample also cover a larger range of luminosities and morphological types than previous studies. 

The models were carried out using the GIPSY software tasks {\sc rotmod} and 
{\sc rotmas}.
MOND fits with a$_{0}$ fixed and free were performed. MOND fits with a$_{0}$ free were needed to re-estimate the average value of a$_{0}$, to identify galaxies in which a$_{0}$ exhibits a significant departure from the standard value of a$_{0}$ = 1.21 $\times$ 10$^{-8}$ cm s$^{-2}$ and to look for any trend between a$_{0}$ and the other parameters of the galaxy.
The MOND fit results are:

\begin{itemize}
  \item An average value of (1.13 $\pm$ 0.50) $\times$ 10$^{-8}$ cm s$^{- 2}$ was measured for the MOND acceleration constant a$_{0}$ which is smaller compared to the standard value of 1.21$\times$ 10$^{-8}$ cm s$^{-2}$ found by \cite{1991MNRAS.249..523B} and should be considered as the standard value since our sample covers a broader range of luminosities and morphological types. 
  \item The RCs predicted by MOND tend to be in good agreement with the observed RCs of bright spirals. 
    \item The difference between the RCs predicted by MOND and the observed RCs is the largest for the following galaxies: DDO 154, IC 2574, NGC 925, NGC 3109, NGC 3198 and NGC 7793 which, with the exception of NGC 3198, tend to be low mass systems. 
  \item A correlation between a$_{0}$ and the extrapolated disk central surface brightness is found. We found that galaxies with higher surface brightnesses require higher values of a$_{0}$ and galaxies with lower surface brightness prefer lower a$_{0}$. This is  problematic for MOND as a new law of physics since a$_0$ should be a constant independent of any galaxy property.
\end{itemize}
Finally, for the mass models with dark matter halos, the ISO halo provides the best fits to the RCs ( \textless$\chi^{2}_{r}$\textgreater = 1.18) compared to MOND with a$_{0}$ fixed (\textless$\chi^{2}_{r}$\textgreater =  9.20) or a$_{0}$ free to vary (\textless$\chi^{2}_{r}$\textgreater =  2.37).
The existence of the scaling relations for the central core of the dark matter halos were investigated for the ISO halo model.
A correlation between the central density $\rho_{0}$ and the core radius \emph{R$_{c}$} was confirmed. This correlation implies that the dark matter halo model could be characterized by only one of the two parameters. 

\section*{Acknowledgments}
We would like to thank Prof. Erwin de Blok for sending us the tabulated RCs and the 3.6 microns surface brightness profiles and also to the THINGS team for making their data publicly available. TR's work was supported by a Square Kilometer Array South Africa (SKA SA) National Astrophysics and Space Science Program (NASSP) bursary. CC's work is based upon research supported by the South African Research Chairs Initiative (SARChI) of the Department of Science and Technology (DST), the SKA SA and the National Research Foundation (NRF).

\appendix
\section*{Notes on Individual Galaxies}
\subsubsection*{DDO 154}
DDO 154 is a  gas dominated nearby dwarf galaxy, classified as an Irregular Barred galaxy or IB(s). A distance of 4.3 Mpc derived from the brightest blue stars by 
\cite{Karachentsev:2004bh} is adopted for this study. This is consistent with the previous results adopted in the literature \citep{1989ApJ...347..760C, 1988ApJ...332L..33C}. The first HI observation of this galaxy was done in November 1985 \citep{1988ApJ...332L..33C}. DDO 154 is part of the THINGS sample which consists of 34 dwarf and spiral galaxies. The kinematics and RCs of 19 galaxies from the THINGS were derived by \cite{de-Blok:2008oq} and these are the highest quality RCs available to date. The inclination and position angle are shown in table 1.
This galaxy has been studied extensively in the context of MOND. 
The mass model results are shown in the first row of Fig. \ref{d1}, the left panel is the model with a DM halo, the middle panel is the MOND model with a$_{0}$ fixed and the right panel for a$_{0}$ free. \cite{1988ApJ...334..130M} consider DDO 154 as "an acute test for MOND" because its internal acceleration is deeply in the MOND regime. The poor MOND fit for DDO 154 has been interpreted as being due to the uncertainties on the measured distance since no measured Cepheid based distance is available. The uncertainties on the inclination is also known to be one of the source of the poor quality MOND fit for DDO 154 due to the unknown thickness of the H{\sc i} disk.
Recently, \cite{2012MNRAS.421.2598A} used a new N-body code which solves the modified Poisson's equation and fits galaxy RCs. They performed four parameters (M/L, stellar \& gas disk scale heights and distance) MOND fits for five galaxies from THINGS. They found that an acceptable MOND fit could be obtained for DDO 154 when the gas disk scale height is taken as z$_{g}$ = 1.5 kpc and with a larger (M/L). 
In our study, a good MOND fit is found for a small value of a$_0 = 0.68 \pm\ 0.02$, nearly half the standard value.
\subsubsection*{NGC 55}
NGC 55 is a barred spiral SB(s)m galaxy member of the Sculptor group. A cepheid distance of 1.9 Mpc was measured by \cite{2008ApJ...672..266G} as part the Araucaria Project. The H{\sc i} RCs was derived by Puche \& Carignan (1991) from VLA observations. MOND produces an acceptable fit for the RC of this galaxy.
\subsubsection*{NGC 247}
NGC 247 is a nearby dwarf galaxy part of the Sculptor group . This galaxy is classified as SB(s)cd. The most recent cepheid distance for NGC 247 is 3.4 Mpc. This was measured as part of the Araucaria Project \cite{2009ApJ...700.1141G}. The 	RC of NGC 247 was taken from \cite{1991ApJ...378..487P}. MOND underestimate the rotation velocities in the inner parts of the RCs which leads to a high reduced chi-squared value for this galaxy. 
\subsubsection*{NGC 300}
NGC 300 is a well known spiral galaxy part of the Sculptor Group classified as SA(s)d. This galaxy has been observed at 21 cm wavelength using the 27.4 m twin-element interferometer of the Owen Valley Radio Observatory \citep{1979ApJ...229..509R} and using the Very large Array with the D and C configuration \citep{1991ApJ...378..487P}. The H{\sc i} RC in \cite{2011MNRAS.410.2217W} derived from H{\sc i} observations using the Australian Compact Array telescope is chosen for this analysis because of its is large radial extend compared to the one done by \cite{1991ApJ...378..487P} derived from VLA data. This is one of the galaxies that MOND could not produce acceptable fit to the observed RCs with a reduced chi-squared of 5.43.

\subsubsection*{IC 2574 and NGC 925}
\cite{Gentile:2011th} mentioned the presence of holes and shells in the HI gas distribution of these two galaxies. The existence of large non-circular motions  have also been noticed by \cite{2008AJ....136.2761O}. Therefore, the poor quality MOND fits for these two galaxies could not be interpreted as a failure for MOND.
However, the RCs we used were derived using the bulk velocity field only \cite{2008AJ....136.2761O}, which takes out most of the effects of local non-circular motions. This is well explained in \cite{de-Blok:2008oq}. However, the presence of a small bar could also introduce large scale distortion in the inner parts, but this is beyond the scope of this work and will therefore be investigated in an upcoming paper using numerical simulation.

\subsubsection*{NGC 2366}
NGC 2366 is a dwarf galaxy member of the M81 group. The RCs derived by \cite{2008AJ....136.2761O} using the bulk velocity field is adopted for this study. MOND produces an acceptable fit to the observed RC of this galaxy.
\subsubsection*{NGC 2403}
This galaxy belongs also to the M81 group of galaxies. It is classified as a barred spiral galaxy SAB(s)cd. NGC 2403 is a well known  bright spiral galaxy. It has been extensively used to test MOND. For example NGC 2403 was part of the sample of \cite{1991MNRAS.249..523B} to estimate the value of the MOND acceleration parameter. The most recent RC
 of NGC 2403 was derived by \cite{de-Blok:2008oq}. NGC 2403 is also part of the sample of \cite{Gentile:2011th} in the context of MOND.
A cepheid distance of 3.22 Mpc is adopted for this work \citep{Freedman:2001eu}. A better fit is obtained with a distance of 3.76 Mpc (see Table 3).
\subsubsection*{NGC 2841}
NGC 2841 is a bright spiral galaxy in the constellation of Ursa Major. Many authors have noticed that MOND cannot reproduce the observed RC of NGC 2841 (eg. \citealt{1991MNRAS.249..523B}). \cite{1996ApJ...473..117S} even considered NGC 2841 as a possible case to falsify MOND saying that a good MOND fit was only possible with very large distance and an unrealistic disk (M/L). In this work,
the discrepancy has largely decreased ($\chi^{2}_{r}$ = 0.896) using the Cepheid distance of 14.1 $\pm$1.5 Mpc \citep{Freedman:2001eu}, and a higher value for a$_{0}$ (cf Table 2) which is consistent with \cite{Gentile:2011th} (see their result with distance constrained). 
\subsubsection*{NGC 3031}
NGC 3031 or M81 is a bright spiral galaxy in the constellation of Ursa Major. NGC 3031 is located at a distance of 3.63 Mpc \cite{Freedman:2001eu}. The existence of non-circular motions is reported by \cite{de-Blok:2008oq}
. For this reason it is excluded from the \cite{Gentile:2011th} sample and the observed RC has not yet been confronted with the MOND formalism. \cite{2008AJ....136.2720T} quantified the non-circular motions for this galaxy and found that they lie between 3 and 15 km s$^{-1}$ for an average of 9 km/s. They also noticed that the outer disk is warped and that there is some disturbance in the velocity field. Allowing a$_{0}$ to vary did not improve the quality of the MOND fit. 
\subsubsection*{NGC 3109}
NGC 3109 is a nearby SB(s)m dwarf galaxy located at a distance of about 1.30 Mpc from us. The first cepheid distance of NGC 3109 was measured by 
\cite{2005ApJ...628..695G} from a total of 19 cepheid variables. 
The first  H{\sc i} RC for this galaxy was derived by 
\cite{1990AJ....100..648J} from a VLA observation using C and D configurations. Another H{\sc i} observation were done using the 64 m Parkes telescope in Australia \citep{2001AJ....122..825B}, but they could not derive the RC because of the lack of spatial resolution. 
\cite{1991MNRAS.249..523B} noticed that the gas component need to be increased by a factor of 1.67 after a comparison with single dish observations in the 21 cm wavelength to be in accord with MOND. 
Recently, \cite{2013AJ....146...48C} obtained new H{\sc i} observations with the Karoo Array Telescope (KAT 7) ( SKA and MeerKAT precursor) in the Karoo desert in South Africa. 
Since the short baselines and the low system temperature make the telescope very sensitive to large scale low surface brightness emission, 
all the H{\sc i} gas is detected by KAT-7. Despite having now the proper gas profile, they conclude that NGC 3109 continues to be problematic for MOND. Since it has the proper gas profile, the RC derived by \cite{2013AJ....146...48C} is used in this study. NGC 3109 still exhibits a large discrepancy between the RC predicted by MOND  and the observed RC. This disagreement between the MOND RC and the observed RC remained even when a$_{0}$ is taken as a free parameter.

 \subsubsection*{NGC 3198}
This is a grand design spiral galaxy in the constellation of Ursa Major. The cepheid distance of 13.80 Mpc of \cite{Freedman:2001eu} is used in this work . NGC 3198 is a well studied galaxy in the context of MOND. Many authors have shown that MOND cannot predict the RC of this galaxy with the standard value of a$_0$, unless adjustment is made on the adopted distance. A much lower distance of 8.6 Mpc is needed to reconcile MOND with the observed RC of NGC 3198 (see: \citealt{Bottema:2002kl}, \citealt{Gentile:2011th}). This is much smaller than the Cepheid distance of 13.8 Mpc, adopted for our study and therefore very unlikely. Despite the non-circular motion induced by the presence of a bar, we can get reasonable DM fits but the discrepancy between the observed RC and the MOND fit is large with a $\chi^{2}_{r}$ = 24.26. MOND overestimates the rotational velocity in the outer parts, which implies more mass. A good MOND fit is  only obtained by letting a$_{0}$ free to vary (see also: \citealt{Gentile:2011th}). 
 
Recently, \cite{2013A&A...554A.125G} derived a new RC for NGC 3198 as part of the HALOGAS (Westerbork Hydrogen Accretion in LOcal GAaxieS) survey \citep{2011A&A...526A.118H}, which aims to study extra-planar gas in the local universe. Their new RC has a larger extent compared to the THINGS RC but with fewer data points in the inner parts. They performed MOND fits by letting the distance free to vary within the uncertainties. They found that MOND can produce a better fit in the outer parts with the new RC but the quality of the fit is much worse compared with those in the literature in the inner parts ( e. g: \citealt{Gentile:2011th}, \citealt{Bottema:2002kl}, \citealt{1991MNRAS.249..523B}). The inner parts of the galaxy contain most of the mass and plays an important role in the mass model. However, it is not well constrained with the new RC because of the lack of spatial resolution. Using our adopted distance, our results suggest that an acceptable MOND fits is only possible if the MOND acceleration constant a$_{0}$ is of about 0.67 x 10$^{-8}$ cm s$^{-2}$, which is about half the standard value. 
\subsubsection*{NGC 3621}
NGC 3621 is classified as SA(s)d galaxy. It is located at a distance of 6.64 Mpc in the constellation of Hydra \citep{Freedman:2001eu}. It is a well behaved galaxy with a flat RC up to very large radii.  Both MOND and dark matter models produce good fits to the observed RC. 
\subsubsection*{NGC 7331}
NGC 7331 is a spiral galaxy in the constellation of Pegasus classified as SA(s)b. A cepheid distance of 14.72 Mpc \citep{Freedman:2001eu} is adopted in this study. The H{\sc i} RC of NGC 7331 have been confronted to MOND by \cite{1991MNRAS.249..523B} using VLA observations and  \cite{Gentile:2011th} using data from the THINGS survey \cite{Walter:2008bs}. Their conclusions that MOND is able to fit the observed RC is confirmed in this work.
\subsubsection*{NGC 7793}
NGC 7793 is a member of the Sculptor Group classified as S(s)d. The most recent cepheid distance for NGC 7793 is 3.43 Mpc. This was measured as part of the Araucaria Project \cite{2005ApJ...628..695G}. The RC derived from the THINGS data is used in this study. The quality of the MOND and dark matter fits are similar for this galaxy (see \cite{Gentile:2011th}
 for the MOND and \cite{de-Blok:2008oq}
 for the dark matter fits). This is confirmed by the chi-square values listed in Table 2 for the MOND and ISO models.


\bsp

\label{lastpage}

\end{document}